\documentclass[prd,aps,onecolumn,preprintnumbers,amsmath,amssymb]{revtex4}%\documentclass[onecolumn,prl,nofootinbib]{revtex4}showpacs,
\usepackage{amsmath}
\usepackage{graphicx}
\usepackage{dcolumn}
\usepackage{bm}
\usepackage{amssymb}
\usepackage{latexsym}

\def\be{\begin{equation}}
\def\ee{\end{equation}}
\def\bea{\begin{eqnarray}}
\def\eea{\end{eqnarray}}

\begin{document}

%\date{\today}

\title{G-Bounce Inflation: Towards Nonsingular Inflation Cosmology with Galileon Field}

\author{Taotao Qiu$^{1,2}$}
\email{qiutt@mail.ccnu.edu.cn}

\author{Yu-Tong Wang$^3$}
\email{wangyutong12@mails.ucas.ac.cn}

\vspace{16mm}

\affiliation{$1$ Institute of Astrophysics, Central China Normal University, Wuhan 430079, China}
\affiliation{$2$ State Key Laboratory of Theoretical Physics, Institute of Theoretical Physics, Chinese Academy of Sciences, Beijing 100190, China}
\affiliation{$3$ School of Physics, University of Chinese Academy of Sciences, Beijing 100049, China \vspace{1mm}}

\pacs{98.80.Cq}
\begin{abstract}

We study a nonsingular bounce inflation model, which can drive the early universe from a contracting phase, bounce into an ordinary inflationary phase, followed by the reheating process. Besides the bounce that avoided the Big-Bang singularity which appears in the standard cosmological scenario, we make use of the Horndesky theory and design the kinetic and potential forms of the lagrangian, so that neither of the two big problems in bouncing cosmology, namely the ghost and the anisotropy problems, will appear. The cosmological perturbations can be generated either in the contracting phase or in the inflationary phase, where in the latter the power spectrum will be scale-invariant and fit the observational data, while in the former the perturbations will have nontrivial features that will be tested by the large scale structure experiments. We also fit our model to the CMB TT power spectrum.

\end{abstract}

\maketitle

\section{Introduction}
The standard model of cosmology regards inflation \cite{Guth:1980zm, Linde:1981mu, Albrecht:1982wi, Starobinsky:1980te, Fang:1980wi, Sato:1980yn} as an important period in the early universe. As a super-fast expansion after the Big Bang, the inflation can solve a series of cosmological problems such as ${\it horizon}$, ${\it flatness}$, ${\it monopole}$ and so on, as well as give rise to scale-invariant scalar perturbations that can fit with the data. However, traditional inflation scenario still needs to be improved, because of the so-called ``singularity problem" which was proposed by Hawking et al. in their early work of singularity theorem \cite{Hawking:1969sw,Borde:1993xh}. According to their proof to this theorem, if we track back inflation to its very beginning, we will generally meet the singularity of the early universe (the Big-Bang singularity). At the singularity, everything blows up and one can not get control of the universe under classical description. Since the singularity occurs before the onset of inflation, it can hardly be solved within inflation scenario itself. This motivates us to find alternative theories in pre-inflation era.

Phenomenologically, there might be quite a few evolutions that can be set in front of inflation in order to get rid of this problem. As an example, the universe may undergo a contracting phase where the scale factor $a(t)$ shrinks initially and then, by some mechanism, ``bounces" into an expanding one \cite{Novello:2008ra}. The whole process can be done non-singularly if at the bouncing point $a_B(t)\neq 0$ \cite{Cai:2007qw}. The bouncing scenario has many interesting properties, for example, the Big-Bang puzzles such as horizon problem and flatness problem can be solved even in contracting phase, and scale-invariant primordial perturbations can be generated, etc \cite{Cai:2007zv}. Moreover, such non-singular scenario can also be non-trivially extended to the cyclic universe \cite{Xiong:2007cn,Piao:2004me}.

In bouncing inflation scenario however, there will be two more latent problems. One of them is the so-called ``ghost instability". This problem claims that field theory models that violates the Null Energy Condition (NEC) will generally give rise to ``ghost" degrees of freedom, which causes the instability. The problems is originated from the ``Ostradski problem" \cite{Woodard:2006nt} in field theory, and the first cosmological application of this problem was on dark energy models \cite{Carroll:2003st} (see also \cite{Onemli:2002hr}, which discussed from quantum aspect). Since a nonsingular bounce violates NEC, normally the ghost will appear. However, recently an interesting ``Galileon" theories \cite{Nicolis:2008in, Deffayet:2009mn} has been shown to be able to avoid such ghost degree of freedom while violating NEC (see pioneering work on such kind of models in \cite{Horndeski:1974wa}). The reason is that although Galileons have higher derivative operators that can violate NEC, due to the delicate design of the Lagrangian, there will only be one dynamical degree of freedom, which can be made normal. The other degrees of freedom is non-dynamical, thus will not lead to any instabilities. Therefore, in this paper we will make use of Galileon theory to build our model. Note that pure Galileon bounce models has also been proposed in the literatures \cite{Qiu:2011cy, Qiu:2013eoa}, which can avoid ghost instability on NEC violation.

The other problem is known as ``anisotropy problem" \cite{Kunze:1999xp}, which will impose additional constraint on the evolution of the universe before the bounce. An exact isotropic universe at the initial time needs somehow fine-tuning, so in realistic models of the early universe, certain amount of anisotropies will exist. In pure inflation scenario, this will not be a problem because the anisotropies will decay in expanding universe, and will eventually be diluted away by inflation to get an isotropic universe. In bounce cosmology however, the anisotropies will grow in contracting phase, and if the growth of the background energy density does not exceed that of anisotropies, the latter will dominate the universe to make it a totally anisotropic one, and probably no bounce will occur and the universe will never enter into an expanding phase, which shows contradiction to our observations. By calculation, we can know that the growth of anisotropies scales as $a^{-6}$, and to have growth of background faster than this speed one must have the equation of state (EoS) larger than unity, which gives constraint to the model building. In fact, a large EoS can be obtained by requiring a negative potential. Though currently such a potential is only for phenomenogically use, it can actually be reduced from more fundamental theories such as Ekpyrotic theories.

The phenomenology of bounce inflation scenario was first studied
in \cite{Piao:2003zm}, where it was shown that such a scenario can
not only obtain scale-invariant scalar perturbations that can fit
with the data, the perturbations generated before the bounce can
also give rise to tilted spectrum, which can explain the
suppression of CMB TT spectrum at large scales. In the second
paper of Ref. \cite{Cai:2007zv}, a bounce inflation model was
realized, but the ghost and anisotropy problems has not been
addressed. In \cite{Qiu:2014nla}, a sketch plot of healthy bounce
inflation model has been drawn with the ways of avoiding those
problems mentioned above, and in \cite{Liu:2013kea}, the 1st-year
PLANCK data has been used to constrain this scenario. In
\cite{Wang:2014abh} (also see \cite{Xia:2014tda}), it was found that the bounce inflation may
generate a large circularly polarized gravitational wave on large
scale, which will lead to the enhancement of TB and EB-mode
correlations on corresponding scale, see \cite{Cai:2015nya} for recent study on large scale power deficits on the CMB fluctuations along with various pre-inflationary models.

In this paper, we aimed at building a realistic bounce inflation model, and fit with the newest PLANCK data. Following \cite{Qiu:2014nla} and also the above arguments, we make use of the Galileon theory the avoid ghosts, and introduce negative potential to keep anisotropies away during the evolution. Moreover, another technical point is how to unify a negative potential with a positive potential which is used to drive a period of inflation, and in the following we will show that, with the help of some ``shape functions", one can naturally match the evolutions of contracting and expanding phases together. In the case of model building the shape functions are shown to be very useful, although its fundamental origination needs to be more explored.

The rest of the paper will be organized as follows: in Sec. II we set up our model, and show by numerics that it can give us non-singular bounce followed by a period of inflation, and the ghost and anisotropy problems will not appear. In Sec. III we calculate the perturbations generated in our model, and show that for perturbations of small wavelengths, nearly scale-invariance will be realized similar to inflation case, while for those of long wavelengths, there will be a blue-tilt which can explain the low-$l$ suppression in CMB TT spectrum. in Sec. IV we will fit our model with the newly released PLANCK data. Sec. V comes as a conclusion.

\section{The Galileon bouncing inflation model}
As has been demonstrated in the introduction, in order to avoid ghost instability, it is useful to make use of Galileon theory to build up the bounce inflation model. The most generalized Lagrangian of Galileon is constructed as \cite{Deffayet:2009mn}:
\bea
{\cal L}&=&\sum_{i=2}^{5}{\cal L}_i~, \nonumber\\
{\cal L}_2&=&K(\phi, X)~,\nonumber\\
{\cal L}_3&=&-G_3(\phi, X)\Box\phi~,\nonumber\\
{\cal L}_4&=&G_{4}(\phi, X)R+G_{4X}\left[
\left(\Box\phi\right)^2-\left(\nabla_\mu\nabla_\nu\phi\right)^2
\right]~,\nonumber\\
{\cal L}_5&=&G_5(\phi, X) G_{\mu\nu}\nabla^\mu\nabla^\nu\phi
-\frac{G_{5,X}}{6}\Bigl[\left(\Box\phi\right)^3
-3\left(\Box\phi\right)\left(\nabla_\mu\nabla_\nu\phi\right)^2
+2\left(\nabla_\mu\nabla_\nu\phi\right)^3
\Bigr]~,\nonumber
\eea
where $X\equiv-\partial_\mu\phi\partial^\mu\phi/2$ is the canonical kinetic term of $\phi$, $\Box\phi\equiv g^{\mu\nu}\nabla_\mu\nabla_\nu\phi$, $R$ is the Ricci scalar and $G_{\mu\nu}\equiv R_{\mu\nu}-g_{\mu\nu}R/2$ is the Einstein tensor. However, in order to make our model simpler, we will only use ${\cal L}_2$ and ${\cal L}_3$ instead of all the ${\cal L}_i$'s. To be explicit, we take the Lagrangian to be:
\be\label{lagrangian}
{\cal L}=k(\phi)X+t(\phi)X^2-V(\phi)-G(X,\phi)\Box\phi~,
\ee
where the last term, which we called the ``G-term", contains the second derivative of the scalar field $\phi$ which will be essential in violating NEC and trigger the bounce. Note that for the kinetic part $k(\phi)X+t(\phi)X^2$, when $k(\phi)=1$, $t(\phi)=0$, it will reduce to trivial kinetic term $X$, and when $k(\phi)=-1$, $t(\phi)=1$ it reduces to the form of ``ghost condensate". We will see in the following that kinetic part of this form is useful in matching bounce with inflation, which was first proposed for ghost-free bounce by Cai et al \cite{Cai:2012va}, and later connected with Supergravity in \cite{Koehn:2013upa}. Following the Lagrangian (\ref{lagrangian}), one can get the equation of motion for $\phi$ as:
\bea
&&[k(\phi)+6t(\phi)X+6G_XH\dot\phi+6HG_{XX}\dot\phi-2(G_\phi+G_{X\phi}X)]\ddot\phi+3H[k(\phi)+2t(\phi)X-2(G_\phi-G_{X\phi}X)]\dot\phi~\nonumber\\
&&+[2k_\phi(\phi)+4t_\phi(\phi)X+6G_X(\dot H+3H^2)-2G_{\phi\phi}]X-k_\phi(\phi)-t_\phi(\phi)X^2+V_\phi(\phi)=0~,
\eea
and the energy density and pressure are
\be
\rho=k(\phi)X+3t(\phi)X^2+3G_XH\dot\phi^3-2G_\phi X+V(\phi)~,~P=k(\phi)X+t(\phi)X^2-2(G_\phi+G_X\ddot\phi)X-V(\phi)~
\ee
respectively.

We can furtherly determine the functions in Lagrangian (\ref{lagrangian}) with the evolutions we want to get. First of all, an anisotropy-free contraction phase which requires the EoS larger than unity generally leads to a negative potential, the simplest Lagrangian of which is:
\be\label{lagrangiancon}
{\cal L}^{con}=X-V^{con}(\phi)~,
\ee
where $V^{con}(\phi)<0$. Meanwhile, the inflationary expansion phase after the bounce will have a positive flat potential:
\be\label{lagrangianinf}
{\cal L}^{inf}=X-V^{inf}(\phi)
\ee
where $V^{inf}(\phi)>0$, $V^{inf}_\phi/(HV^{inf})\ll 1$. Therefore, we will choose functions $k(\phi)$, $t(\phi)$, $G(X,\phi)$ and $V(\phi)$ so that they can approach Lagrangians (\ref{lagrangiancon}) and (\ref{lagrangianinf}) in the limits of past and future, respectively. As a simple example, we choose forms of $k(\phi)$, $t(\phi)$ and $G(X,\phi)$ as:
\be\label{function}
k(\phi)=1-\frac{2k_0}{[1+2\kappa_1(\phi/M_p)^2]^2}~,~t(\phi)=\frac{1}{M_p^4}\frac{t_0}{[1+2\kappa_2(\phi/M_p)^2]^2}~,~G(X,\phi)=\frac{1}{M_p^3}\frac{\gamma X}{[1+2\kappa_2(\phi/M_p)^2]^2}~.
\ee
and the potential is:
\be\label{potential}
V(\phi)=[1-\tanh(\lambda_1\frac{\phi}{M_p})]V^{con}(\phi)+[1+\tanh(\lambda_2\frac{\phi}{M_p})]V^{inf}(\phi)~,~V^{con}(\phi)=-V_0 e^{c\phi/M_p}~,V^{inf}(\phi)=\Lambda^4(1-\frac{\phi^2}{v^2})^2~,
\ee
with $k_0$, $t_0$, $\gamma$, $\kappa_1$, $\kappa_2$, $\lambda_1$, $\lambda_2$, $V_0$, $c$, $\Lambda$ and $v$ are all free parameters. Note that $V^{con}(\phi)$ is inspired from the Ekpyrotic potential \cite{Khoury:2001wf, Qiu:2013eoa, Fertig:2013kwa}, and $V^{inf}(\phi)$ is the well-known spontaneously symmetry breaking potential \cite{Olive:1989nu} which fits well with the PLANCK data \cite{Ade:2013uln}. So the choice of this potential is not only simple, but may have a strong background of fundamental theories.

From Eqs. (\ref{function}) and (\ref{potential}) one can see that, in the $(\phi/M_p)\gg 1$ limit, we have
\be\label{contracting}
k(\phi)=1~,~t(\phi)=G(X,\phi)=0~,~V(\phi)=V^{con}(\phi)~,
\ee
while in the $(\phi/M_p)\ll -1$ limit, we have:
\be\label{expanding}
k(\phi)=1~,~t(\phi)=G(X,\phi)=0~,~V(\phi)=V^{inf}(\phi)~,
\ee
which can give us (\ref{lagrangiancon}) and (\ref{lagrangianinf}) respectively. Around $\phi=0$ however, $t(\phi)$, $G(X,\phi)$ and the second term of $k(\phi)$ becomes large, so the linear term of $X$ will flip its sign, while the $X^2$-term and G-term will be dominate. Then for the solution where $\phi$ increases monotonically (which is a very natural solution), one will get the anisotropy-free contraction in the past and inflation in the future, between which the bounce is triggered around $\phi=0$.

\begin{figure}[htbp]
\centering
\includegraphics[scale=0.45]{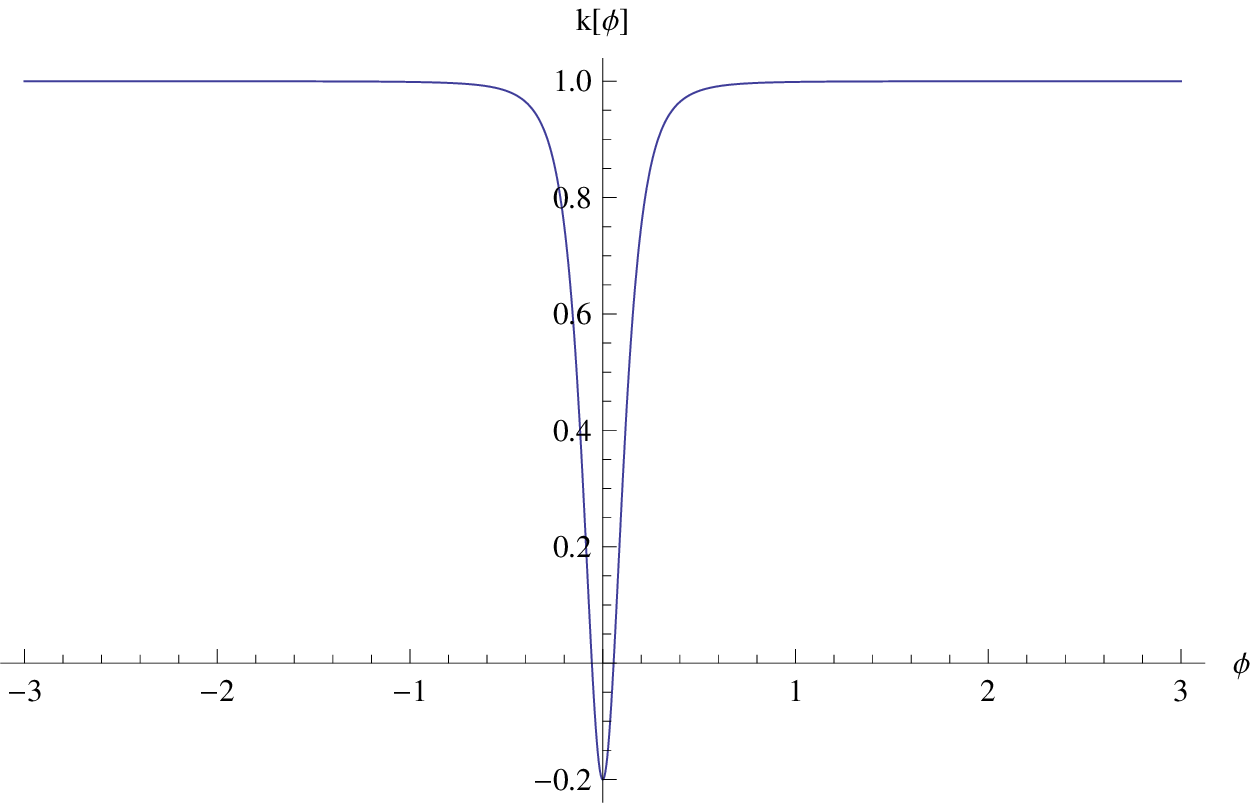}
\includegraphics[scale=0.45]{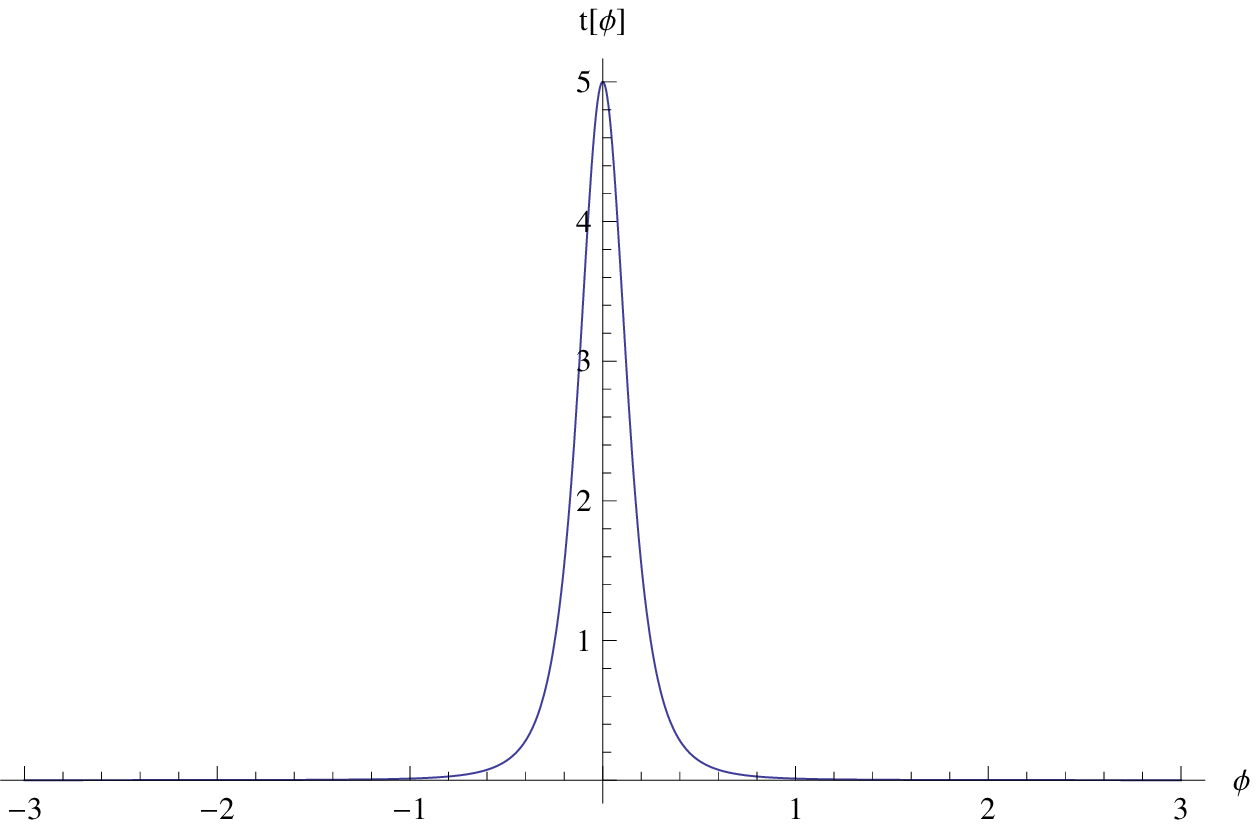}
\includegraphics[scale=0.45]{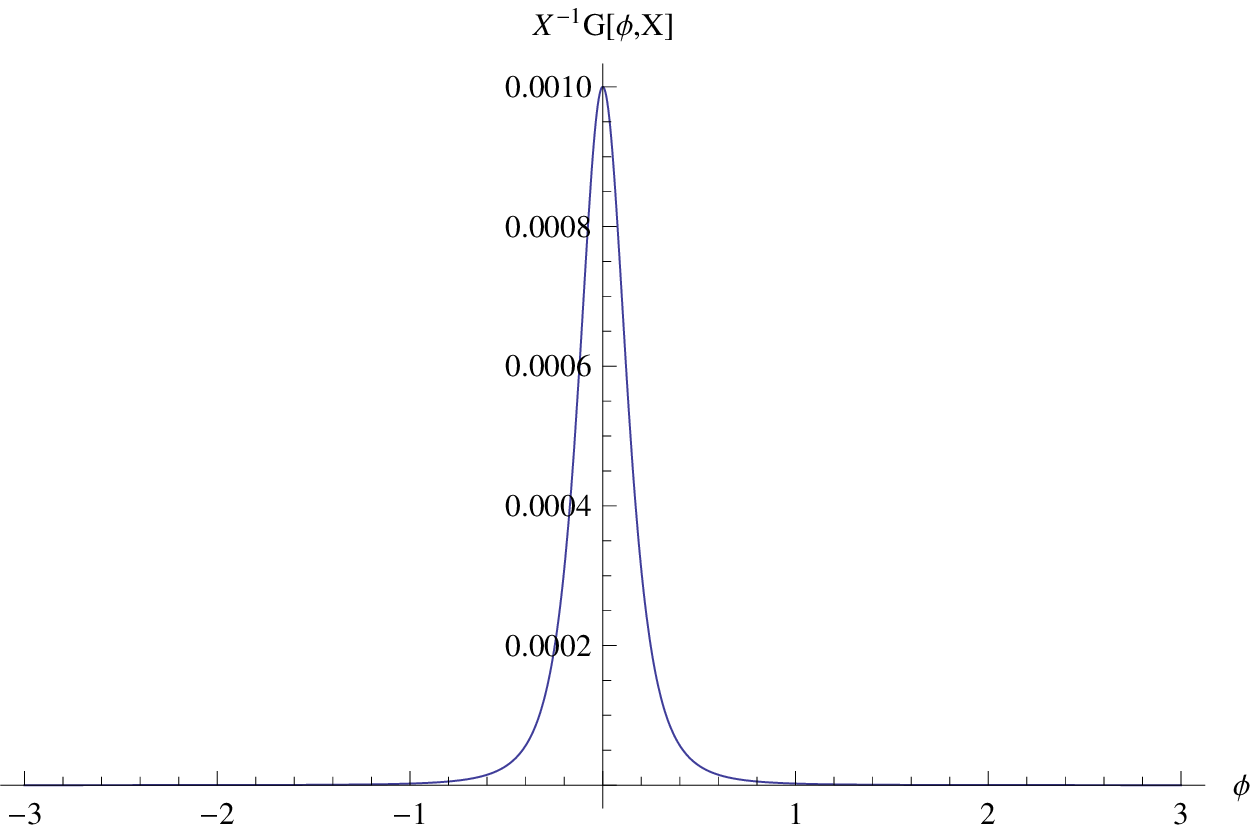}
\caption{Plots of functions $k(\phi)$, $t(\phi)$ and $X^{-1}G(X,\phi)$ in Eq. (\ref{function}) w.r.t. the scalar, $\phi$. Note that the last one is also function of $\phi$ only. The parameters are chosen as $k_0=0.6$, $\kappa_1=15$, $t_0=5$, $\kappa_2=10$, $\gamma=1\times10^3$. In such a choice, all the three functions have nontrivial value only around the bouncing point, which is useful to trigger the bounce. }\label{shapefunction}
\end{figure}

Furthermore, we would like to emphasize that by having the shapes of $k(\phi)$, $t(\phi)$ and $G(X,\phi)$, one can successfully control each term such that the higher derivative term only dominates for a while in order to trigger the bounce, while disappears before and after that, not disturbing the evolution before and after the bounce. In fact, if we do not control the higher derivative term, it will usually become more and more important after dominance, leading to a very fast evolution of the universe instead of a modest one, or even the ``big-rip" singularity. Moreover, using $\tanh$-like shape function, we can naturally connect $V^{con}(\phi)$ and $V^{inf}(\phi)$ so that the field $\phi$ can evolve naturally from a lower potential to a higher one, leading to inflation. Therefore we can see that, shape functions are actually very useful in constructing models. We also show the shape functions and potential explicitly by numerics in Fig. \ref{shapefunction} and \ref{potential}. For pioneer works of using this shape functions to model building, see \cite{Cai:2012va, Koehn:2013upa} for ghost-free bounce and see \cite{Zhang:2006ck} for dark energy models.

\begin{figure}[htbp]
\centering
\includegraphics[scale=0.6]{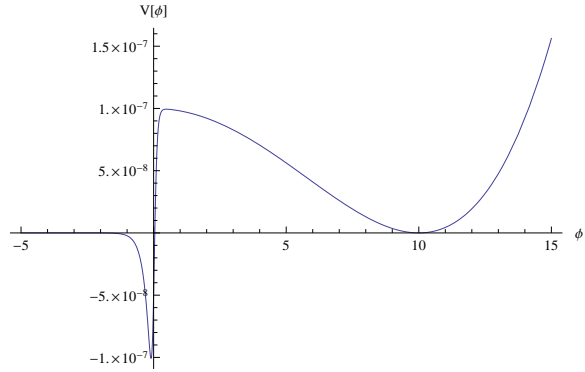}
\caption{Plots of potential $V(\phi)$ in Eq. (\ref{potential}) w.r.t. the scalar, $\phi$. The parameters are chosen as $\lambda_1=\lambda_2=10$, $V_0=0.7M_p^4$, $c=\sqrt{20}$, $\Lambda\approx1.5\times10^{-2}M_p$, $v=10M_p$. From the plot we can see that the shape of the potential shares the same shape of Ekpyrotic potential $V^{con}(\phi)$ for negative $\phi$ while that of symmetry breaking inflaton potential $V^{inf}(\phi)$ for positive $\phi$. We will see below that $\phi=0$ is almost the division of contracting and expanding phases.}\label{potential}
\end{figure}

Now let's go to a little bit more detail about the field evolution both before and after the bounce, which actually become very simple. In the contracting phase before the bounce when Eq. (\ref{contracting}) applies, the equation of motion of $\phi$ becomes:
\be\label{eomcontracting}
\ddot\phi+3H\dot\phi-\frac{c}{M_p}V_0e^{c\phi/M_p}=0~.
\ee
%The scaling solution: \be\label{scaling}
%\phi(t)\simeq-\frac{2}{c}\ln(t_\ast-t)~,~~~\dot\phi(t)\simeq\frac{2}{c(t_\ast-t)}~
%\ee
%requires the relation $V_0=2(1-6/c^2)/c^2$, where $t_\ast$ is some timescale.
The energy density and pressure of $\phi$ thus becomes
%\be
%\rho^{con}\simeq X+V^{con}=\frac{12}{c^4(t_\ast-t)^2}~,~p^{con}\simeq X-V^{con}=\frac{4}{c^2}(1-\frac{3}{c^2})\frac{1}{(t_\ast-t)^2}~,
%\ee
\be
\rho^{con}\simeq X+V^{con}=\frac{1}{2}\dot\phi^2-V_0e^{c\phi/M_p}~,~p^{con}\simeq X-V^{con}=\frac{1}{2}\dot\phi^2+V_0e^{c\phi/M_p}~,
\ee
while the EoS and the slow-roll parameter of the universe are
\be\label{eoscontracting}
w_c\equiv\frac{P^{con}}{\rho^{con}}\simeq1+\frac{4V_0e^{c\phi/M_p}}{\dot\phi^2-2V_0e^{c\phi/M_p}}~,~~~\epsilon_c\equiv\frac{3}{2}(1+w_c)=3+\frac{6V_0e^{c\phi/M_p}}{\dot\phi^2-2V_0e^{c\phi/M_p}}~.
\ee
Since $\dot\phi^2>2V_0e^{c\phi/M_p}$ due to the requirement of positivity of the energy density, $w_c$ is obviously larger than unity, and the growth of the energy density $\rho\propto a^{-3(1+w_c)}$ will be larger than the anisotropy, avoiding the dominance of the latter. So the negative potential is rather useful in avoiding the cosmic anisotropy problem.

\begin{figure}[htbp]
\centering
\includegraphics[scale=0.45]{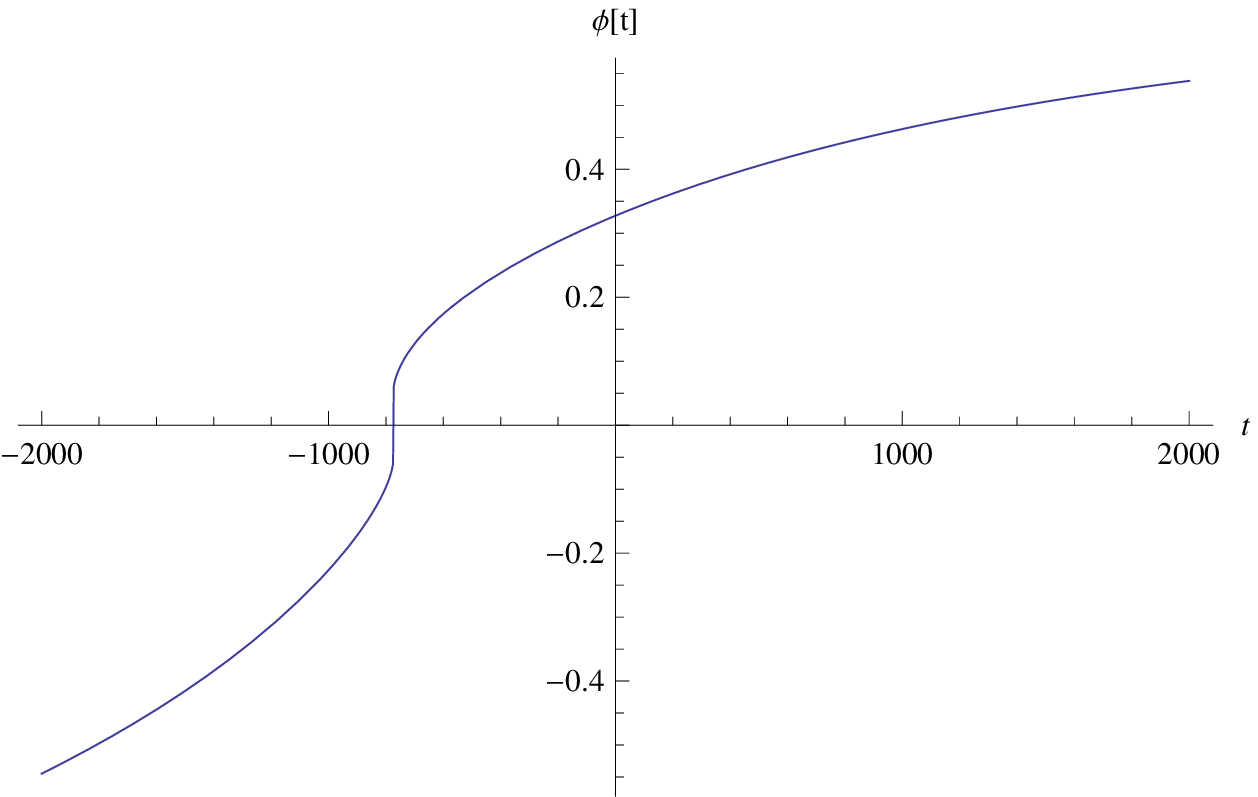}
\includegraphics[scale=0.45]{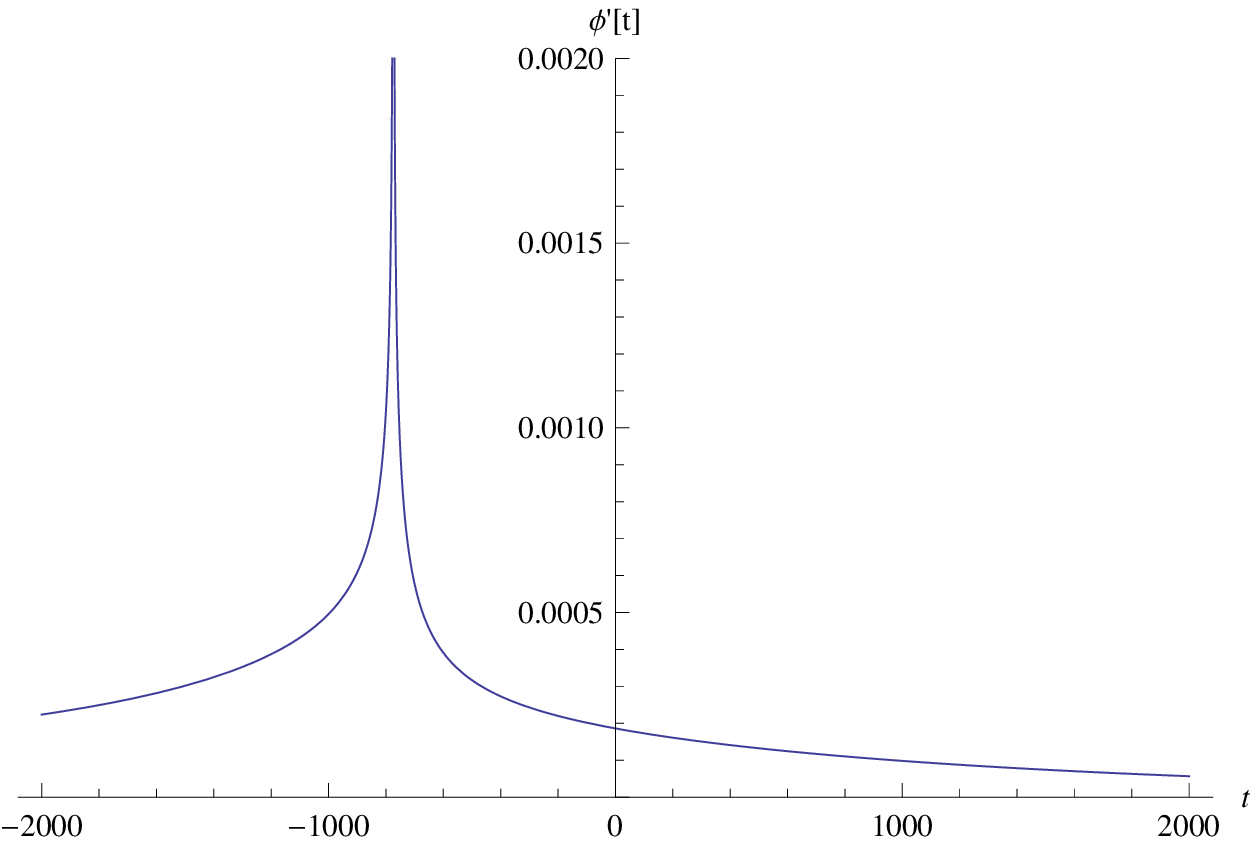}\\
\includegraphics[scale=0.45]{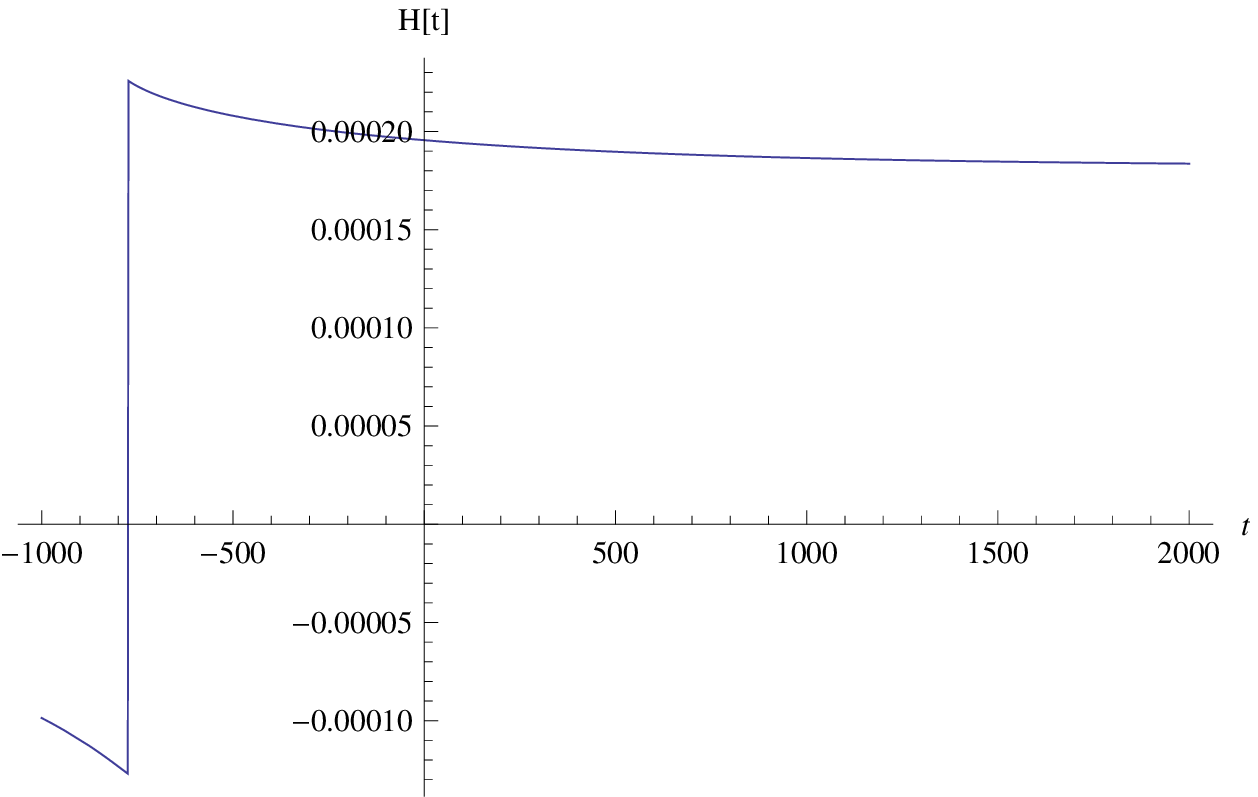}
\includegraphics[scale=0.45]{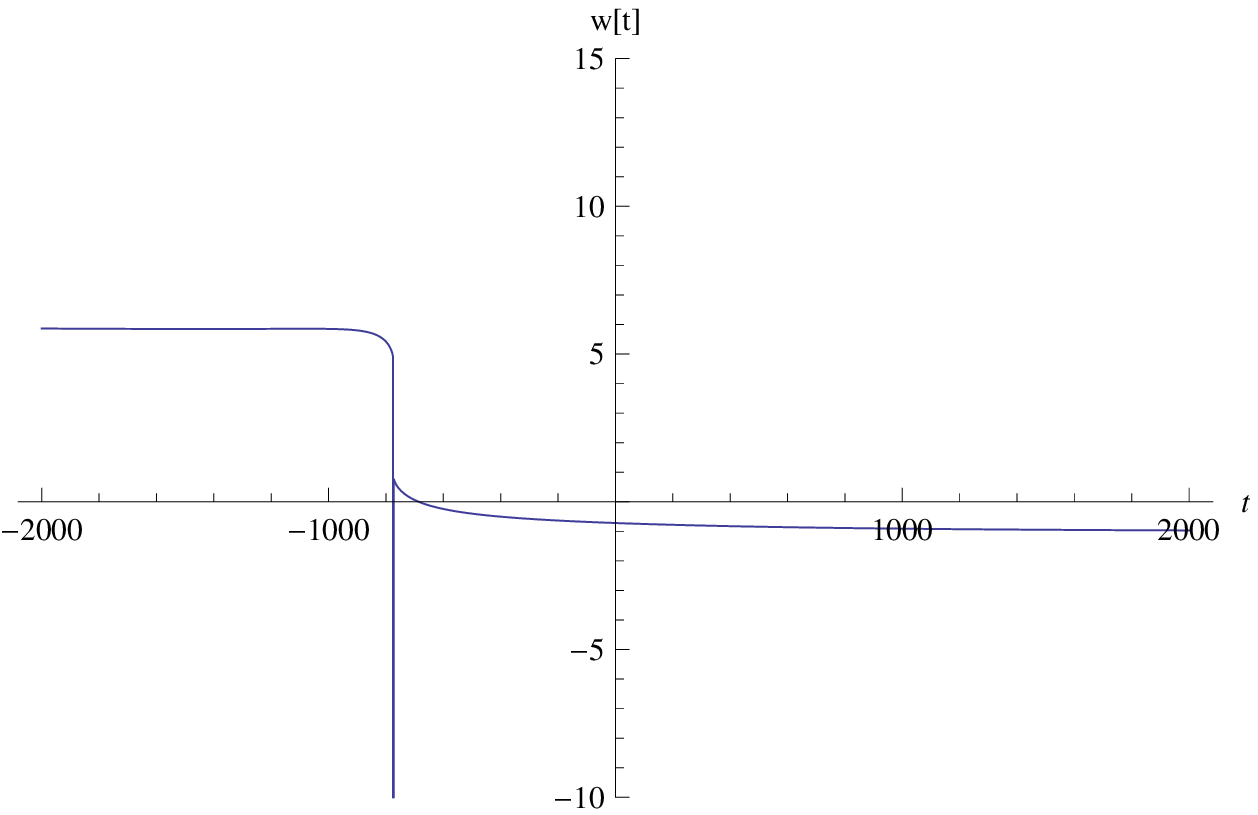}
\includegraphics[scale=0.45]{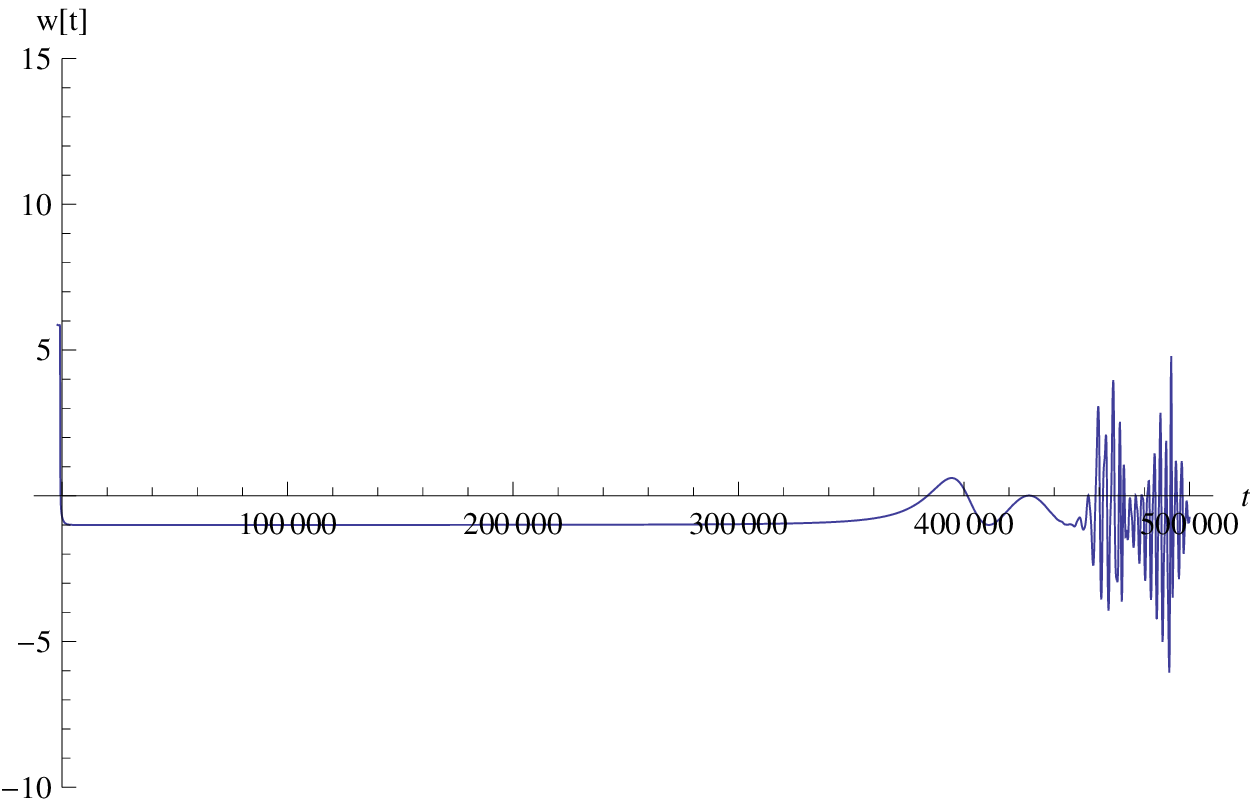}
\caption{Top row: Plots of scalar $\phi$ and its velocity $\dot\phi$ w.r.t. cosmic time $t$. During the evolution, $\phi$ goes from negative to positive values, and as can be seen below, the switching point is nearly the bouncing point. The velocity peaks at the bouncing point, and decay towards both past and future directions. Bottom row: Plot of Hubble parameter $H$ (left) and EoS $w$ (middle and right). The hubble parameter transits from negative to positive values at the bouncing point, and will approach to certain constant positive value, indicating a de-Sitter (inflationary) expansion after the bounce. The middle plot is the EoS near the bounce region. From the plot we see that while before bounce it approach to value larger than unity, it goes down to minus infinity at the bounce point and approaches to $-1$ after the bounce. The right plot is the EoS during whole evolution, and since the inflationary period is very long, the features near the bounce is not viewable. Nevertheless, we can see that after inflation, EoS shows some oscillating behavior, referring to the reheating process.}\label{background}
\end{figure}

Setting the initial values of $\phi$ be negative, and once $\phi$ approaches 0 with certain amount of velocity $\dot\phi$, the higher derivative terms is turned on, which is responsible to trigger the bounce. After bounce however, as Eq. (\ref{expanding}) applies, these terms decay away again, and one will have:
\be
\ddot\phi+3H\dot\phi+2\Lambda^4\frac{\phi}{v^2}(\frac{\phi^2}{v^2}-1)=0~,
\ee
which gives the solution of inflation. According to the potential, one can express the slow-roll parameter $\epsilon_e$ as well as the e-folding number $N_e$ of the inflation as:
\be\label{epsilon}
\epsilon_e(\phi)\simeq\frac{M_p^2}{2}\left(\frac{V_\phi}{V}\right)^2=\frac{8M_p^2\phi^2}{(\phi^2-v^2)^2}~,~ N_e\simeq\int^{\phi_f}_{\phi_i}\left(\frac{V}{V_\phi}\right)d\phi=\left(\frac{\phi^2}{8}-\frac{v^2}{4}\ln\phi\right)\bigg|^{\phi_f}_{\phi_i}~,
\ee
By requiring $\epsilon_e(\phi_f)=1$ and $N_e(\phi)=60$, the field values at starting and ending points of inflation can be calculated as:
\be
\phi_i\approx 1.9M_p~,~\phi_f\approx 11.66M_p~.
\ee
One can see again that the $\phi$ has monotonically increasing evolution. The potential has valley with a minimum at $v=13M_p$, and when $\phi$ falls down the valley, it will become oscillating rapidly, which reheats our universe. The whole evolution of the $\phi$, $\dot\phi$, $H$ and $w$ are sketched in Fig. \ref{background}.

\section{perturbations}
\subsection{scalar perturbations}
\begin{figure}[htbp]
\centering
\includegraphics[scale=0.6]{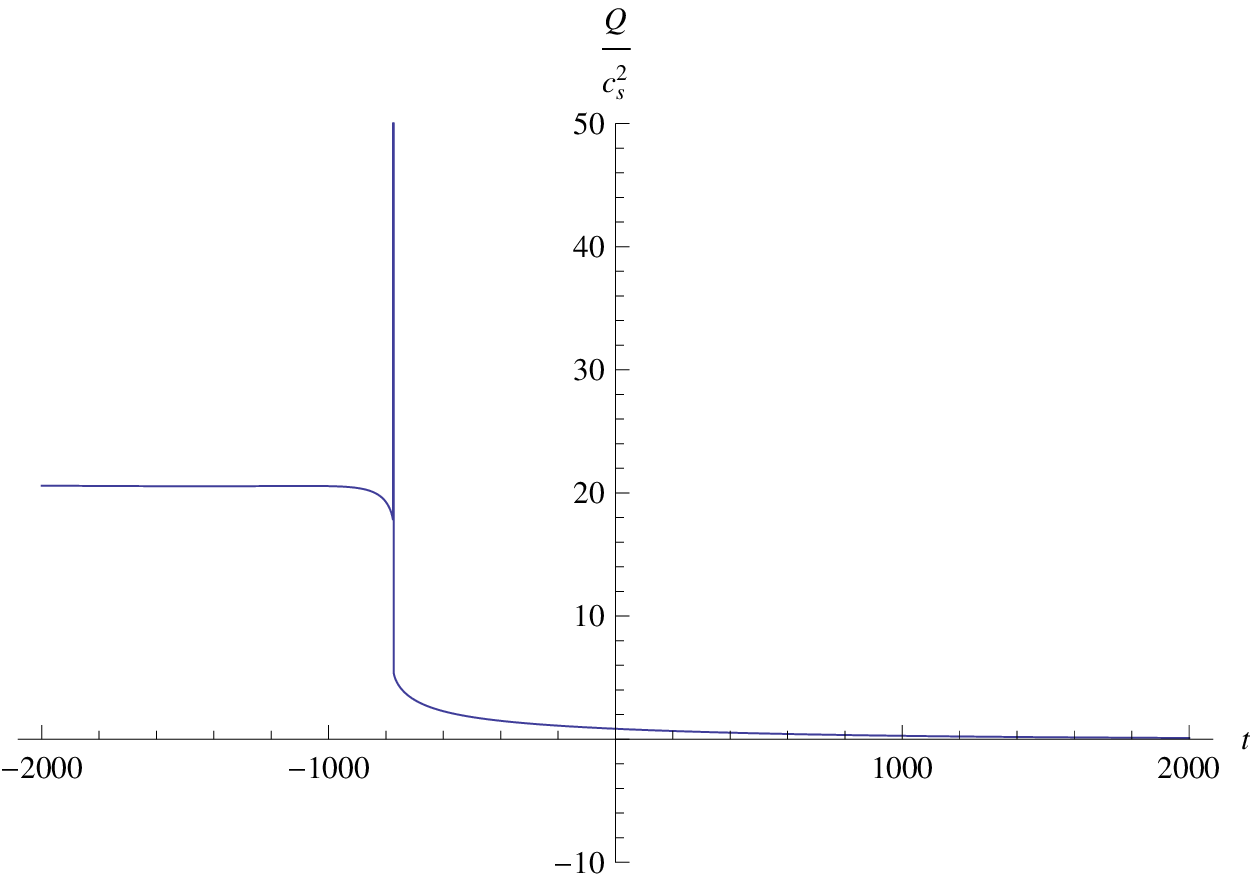}
\includegraphics[scale=0.6]{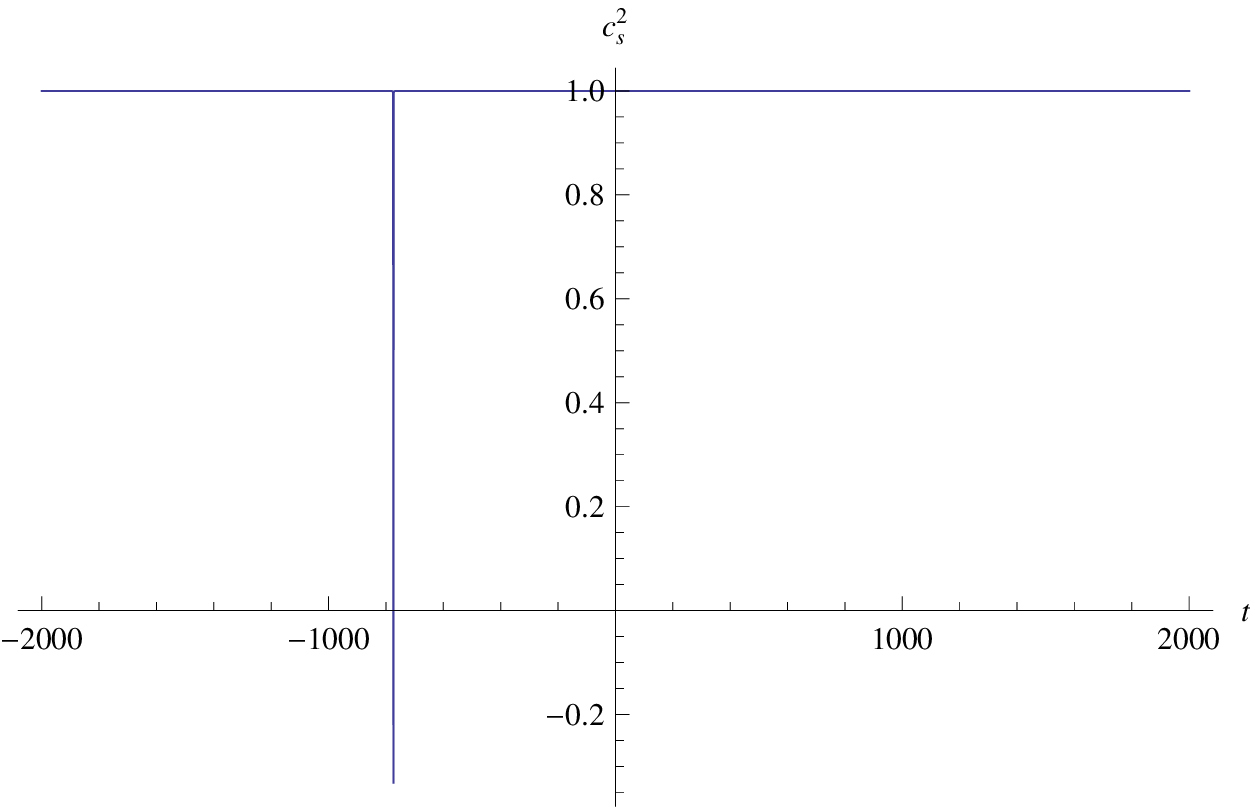}
\caption{Plots of functions $Q/c_s^2$ and $c_s^2$ during the whole evolution. The parameters are chosen as before. We can see that during the whole evolution $Q/c_s^2>0$, indicating that we don't have any ghost instability in our model. In regions away from the bounce, $Q/c_s^2$ reduces to constant proportional to the slow-roll parameter. $c_s^2$ is trivially equal to unity in regions away from the bounce, however around bouncing point, it goes down to below zero, causing short period of gradient instability.}\label{stability}
\end{figure}

In this section, we analyze the perturbations that could be generated in this model. Following the general formulae for calculating perturbations of Galileon models in \cite{Kobayashi:2010cm, DeFelice:2011uc}, the 2-nd order perturbed lagrangian based on (\ref{lagrangian}) turns out to be:
\be\label{pertlagrangian}
{\cal S}^{(2)}=\frac{1}{2}\int d\eta d^3xa^2\frac{Q}{c_{s}^2}\Bigl[\zeta^{\prime 2}-c_{s}^2(\partial\zeta)^2\Bigr]~,
\ee
where
\bea\label{Q}
Q&=&\frac{2M_p^4X}{(M_p^2H-G_XX\dot\phi)^2}[k(\phi)+2t(\phi)X+2(G_X+G_{XX}X)\ddot\phi+4HG_X\dot\phi-\frac{2G_X^2X^2}{M_p^2}]~,\\
\label{cs2}
c_s^2&=&\frac{(M_p^2H-G_XX\dot\phi)^2}{2M_p^4X}[k(\phi)+6t(\phi)X+6H(G_X+G_{XX}X)\dot\phi+\frac{6G_X^2X^2}{M_p^2}]^{-1}Q~.
\eea
Here $\zeta$ is the variable of adiabatic perturbation, while $c_s^2$ is the sound speed squared of $\zeta$, representing how fast the perturbations can propagate. From the quantum field theories, the ghost-free condition of $\zeta$ is $Q/c_s^2>0$. Moreover, the gradient stability requires $0\leq c_s^2\leq 1$. From Eqs. (\ref{Q}) and (\ref{cs2}) one can see that, for most of the evolution when approximations (\ref{contracting}) and (\ref{expanding}) are reached, we have $Q\simeq2M_p^2\epsilon_c$ or $2M_p^2\epsilon_e$ and $c_s^2\simeq 1$. However, for the region near the bounce, evolution are very complicated, and numerical calculations are required. Here we numerically plot $Q/c_s^2$ and $c_s^2$ in Fig. \ref{stability}.

\begin{figure}[htbp]
\centering
\includegraphics[scale=0.3]{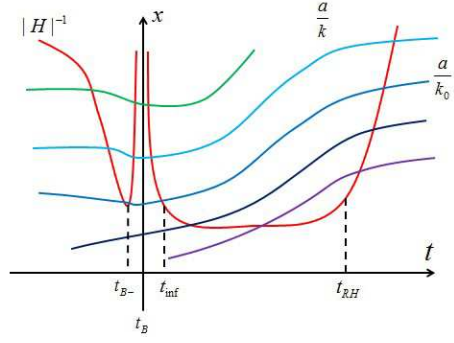}
\caption{The sketch plot of the horizon of our model (red) and the fluctuation modes that cross the horizon before or during inflation period (from green to purple). Since before the bounce the universe evolves slowly with a small value of Hubble parameter, the Hubble horizon is large at most of the time, with a sharp damp near the bounce. However, when bounce happens, it goes to infinity as $H=0$. At the inflation era when $H$ becomes nearly a constant, the horizon also behaves as a constant. The fluctuations with $k\leq k_0$ exit the horizon at contracting phase, thus will stay outside the horizon all over the inflation region. Those with $k>k_0$ exits the horizon at inflationary era, and will stay inside the horizon till the universe expands, and exit the horizon at somewhere during the inflation era.}\label{horizon}
\end{figure}

From the Fig. \ref{stability} we can see that, in the whole evolution (including the bounce) we have positive $Q/c_s^2$, indicating that our model is completely ghost-free. However, during the bounce region, the numerics show that there is a instant period of negative $c_s^2$, which may induce fast increase of the perturbations. We will later show what sequent it will cause to have negative $c_s^2$ and how it can be ameliorated. In fact, the same issue also appears in the model given by \cite{Cai:2012va, Koehn:2013upa}. Let's first write down the equation of motion for $\zeta$ according to lagrangian (\ref{pertlagrangian}):
\be\label{eompertscalar}
u^{\prime\prime}+c_s^2\nabla^2u-\frac{z^{\prime\prime}}{z}u=0~,
\ee
where
\be
u\equiv z\zeta~,~z\equiv\frac{a\sqrt{Q}}{c_s}~,
\ee
and ``$\prime$" and ``$\nabla$" denotes derivatives with respect to conformal time $\eta\equiv\int a^{-1}(t)dt$ and space $x$ respectively. When transformed into momentum space, one has $u(x,t)\rightarrow u_k(t)$ for each wavenumber $k$, and $\nabla^2$ can be substituted into $-k^2$. According to the numerical result in the last section, we can plot the evolutions of fluctuation modes for different wavelengths $\lambda_{phys}=a/k$ as well as the Hubble horizon of our model as in Fig. \ref{horizon}. From the sketch plot one can see that, these fluctuation modes can been divided into two parts, one is of small $k$'s (green, cyan and blue), while the other is of large $k$'s (navy and purple). The former modes are generated in the contracting phase, and will exit the horizon before bouncing, while the latter ones are generated in expanding phase, and will exit the horizon after bouncing. As will be shown very soon, they will have very different evolution and $k$-dependence. We remark the dividing mode between the two as $k_0$ (the blue one), the wavelength of which is just equal to horizon at the beginning of the bounce region.
\subsubsection{initial condition}
Since during the whole evolution, the background is different at each stage, and will lead to different evolution of perturbation. In the following subsections, we will analyze evolution in each stage separately, and joint them together using matching conditions to get the final power spectrum.

At far past, every mode is inside the horizon. In this region, we solve the Eq. (\ref{eompertscalar}) and taking the $k^2\gg z^{\prime\prime}/z$ limit, and obtain the solution:
\be
u_k=Ae^{ik\int_{-\infty}^\eta c_s d\eta^\prime}~.
\ee
As we know that in contracting phase except near the bouncing phase, $Q$ and $c_s$ can be approximated as $1$. Moreover, the coefficient $A$ can be normalized by requiring a quantum origin of the perturbations \cite{Mukhanov:1990me}. We assume that the perturbations starts from Bunch-Davies vacuum as usual, and from the quantum condition, we have $A=(2k)^{-1/2}$. So the initial condition for $u$ is:
\be\label{ic}
u_k=\frac{1}{\sqrt{2k}}e^{i k\eta}~.
\ee
\subsubsection{contracting phase}
In the most part of contracting phase, the model will be governed by the potential term while the higher derivative terms are severely suppressed. In this case, the approximation (\ref{contracting}) applies, and from (\ref{Q}) and (\ref{cs2}) one has $Q\simeq2M_p^2\epsilon_c$ and $c_s^2\simeq 1$. Moreover, if the universe evolves according to the scaling solution of Eq. (\ref{eomcontracting}) with the EoS being as (\ref{eoscontracting}), the scale factor of the universe in contracting phase can be parameterized as
\be
a^{con}(\eta)\sim(\eta-\tilde\eta_{B-})^{\frac{1}{\epsilon_c-1}}~,
\ee
where $\tilde\eta_{B-}\equiv\eta_{B-}-[(\epsilon_c-1){\cal H}_{B-}]^{-1}$, and $\eta_{B-}$ and ${\cal H}_{B-}$ are the conformal time and conformal Hubble parameter at the end point of the contracting phase, respectively. The perturbation equation (\ref{eompertscalar}) then becomes
\be
u_k^{\prime\prime}+k^2u-\frac{2-\epsilon_c}{(\epsilon_c-1)^2}\frac{u}{(\eta-\tilde\eta_{B-})^2}=0~.
\ee
The solution of the above equation is Hankal function:
\be\label{solcontracting}
u_k=\sqrt{|\eta-\tilde\eta_{B-}|}[c_1(k)J_{\nu_-}(k|\eta-\tilde\eta_{B-}|)+c_2(k)J_{-\nu_-}(k|\eta-\tilde\eta_{B-}|)]~,
\ee
with the coefficients $c_1(k)$ and $c_2(k)$ to be determined by the initial conditions (\ref{ic}). Here
\be\label{numinus}
\nu_-=\frac{3-\epsilon_c}{2(\epsilon_c-1)}~.
\ee
In the region far away from the bounce where all the fluctuation modes are deep inside the horizon, one have $k^2\gg |z^{\prime\prime}/z|$, and the solution (\ref{solcontracting}) can be approximated as:
\bea\label{solcontractingsub}
u_k&\simeq&c_1(k)\sqrt{\frac{2}{\pi k}}\cos(k|\eta-\tilde\eta_{B-}|)~,~c_2(k)\sqrt{\frac{2}{\pi k}}\sin(k|\eta-\tilde\eta_{B-}|)~,\nonumber\\
\zeta_k&\simeq&\frac{c_1(k)}{a_{B-}}\sqrt{\frac{1}{\pi\epsilon_c k}}\left(\frac{\eta-\tilde\eta_{B-}}{\eta_{B-}-\tilde\eta_{B-}}\right)^{\frac{1}{1-\epsilon_c}}\cos(k|\eta-\tilde\eta_{B-}|)~,~\frac{c_2(k)}{a_{B-}}\sqrt{\frac{1}{\pi\epsilon_c k}}\left(\frac{\eta-\tilde\eta_{B-}}{\eta_{B-}-\tilde\eta_{B-}}\right)^{\frac{1}{1-\epsilon_c}}\sin(k|\eta-\tilde\eta_{B-}|)~,
\eea
In the region near the bounce, the horizon shrinks rapidly to a minimum value. In this case, some fluctuation modes with small $k$, namely $k^2\ll |z^{\prime\prime}/z|$ will exit the horizon and become superhorizon modes in the contracting phase. The approximate behavior of these modes according to the solution (\ref{solcontracting}) will be:
\bea\label{solcontractingsup}
u_k&\simeq&\frac{c_1(k)}{2^{\nu_-}\Gamma(\nu_-+1)\sqrt{k}}(k|\eta-\tilde\eta_{B-}|)^\frac{1}{\epsilon_c-1}~,~\frac{c_2(k)}{2^{-\nu_-}\Gamma(-\nu_-+1)\sqrt{k}}(k|\eta-\tilde\eta_{B-}|)^\frac{\epsilon_c-2}{\epsilon_c-1}~,\nonumber\\
\zeta_k&\simeq& \frac{c_1(k)[(\epsilon_c-1){\cal H}_{B-}]^\frac{1}{1-\epsilon_c}}{2^{\nu_-}\Gamma(\nu_-+1)\sqrt{2\epsilon_c}a_{B-}}k^{-\frac{\epsilon_c-3}{2(\epsilon_c-1)}}~,~\frac{c_2(k)[(\epsilon_c-1){\cal H}_{B-}]^\frac{1}{1-\epsilon_c}}{2^{-\nu_-}\Gamma(-\nu_-+1)\sqrt{2\epsilon_c}a_{B-}}k^{-\frac{\epsilon_c-3}{2(\epsilon_c-1)}}(k|\eta-\tilde\eta_{B-}|)^\frac{\epsilon_c-3}{\epsilon_c-1}~,
\eea
while the large $k$ mode still remains the same as (\ref{solcontractingsub}).

The subhorizon solution in the contracting phase, which describes the modes deep inside the horizon, can be connected to the initial condition (\ref{ic}) to find out the coefficients $c_1(k)$ and $c_2(k)$. Using (\ref{ic}) and (\ref{solcontractingsub}) we will get:
\be\label{c1c2}
c_1(k)=\frac{\sqrt{\pi}}{2}~,~c_2(k)=i\frac{\sqrt{\pi}}{2}~.
\ee

One can see that while subhorizon modes shows oscillating behavior, the superhorizon modes splits into two branches, one keeps constant and the other is decaying. Assuming the two modes are generated equally, one could easily understood that for modes that exits horizon at contracting phase, the constant mode will dominate over the decaying one, and will be inherited by the perturbations through the bounce and in expanding phase. We will investigate the following evolution of perturbations in the coming sections.
\subsubsection{through the bounce}
After the Hubble parameter of the universe reaches the minimum
value, it will no longer evolve as an monotonic function which
keeps negative, but rather turns around towards positive values.
This will cause the universe contract more and more slowly, while
finally it stops contracting and becomes expanding, triggering the
bounce. The Hubble parameter will go from its minimum value to its
maximum value, the process during which we call as a bouncing
phase. During the bouncing phase, the Hubble parameter is
increasing rather than decreasing, and shape functions such as
$k(\phi)$, $t(\phi)$ and $G(X,\phi)$, as well as $Q$ and $c_s^2$,
will have nontrivial features, which will make Eq.
(\ref{eompertscalar}) hard to solve. For sake of analyticity, we
assume that the Hubble parameter increases as: \be H\simeq
\alpha(t-t_B)~, \ee when $t_B$ is the time point when the bounce
happens. Then we have \be\label{abounce} a\simeq a_B
e^{\alpha(t-t_B)^2/2}\simeq a_B[1+\frac{1}{2}\alpha(t-t_B)^2]~ \ee
for the scale factor. Here $a_B$ refers to the scale factor at
$t_B$. From the numerical calculation, it can be identified that
around the bouncing point the first two terms of the Lagrangian
(\ref{lagrangian}) will be dominant, so one roughly have:
\be\label{qandcs2t}
Q\simeq\frac{2M_p^4k(\phi)X}{(M_p^2H-G_XX\dot\phi)^2}~,~c_s^2\simeq\frac{k(\phi)}{6t(\phi)X}~,
\ee respectively. Using Friedmann equations, one can furtherly
simplify $c_s^2$ to be \be
c_s^2\simeq-\frac{1}{2}\frac{1+2\alpha(t-t_B)^2}{1+3\alpha(t-t_B)^2}~,
\ee and near the bouncing point where the $t$-dependent terms can
be viewed as higher order. Moreover, near the bouncing point, the
hubble parameter will be small compared to the term $G_XX\dot\phi$
in the dominator of $Q$ (at bouncing point $H=0$), where $G_X$ is
obtained from $G(X,\phi)$ in (\ref{function}). From
(\ref{function}), one can notice that in our model
$G_X=M_p(\gamma/t_0)t(\phi)$, so $Q$ can be expressed as \be
Q\simeq\frac{2M_p^4k(\phi)X}{G_X^2X^2\dot\phi^2}\simeq\frac{M_p^2t_0^2k(\phi)X}{\gamma^2t(\phi)^2X^3}\simeq-\frac{3t_0^2X}{\gamma^2\alpha}~.
\ee

It is convenient to approximately parameterize X as function of
$t$ \cite{Cai:2012va}: \be X\simeq X_B e^{-\chi(t-t_B)^2}~, \ee
where $\chi$ is the parameter and $X_B$ is the value of $X$ at the
bouncing point. From the numerical calculation we find that the
parametrization can efficiently mimic the evolution of $X$ near
the bouncing point.

It is convenient to transfer everything into conformal coordinate
where the time is denoted by the conformal time $\eta$. From the
definition of $\eta$ one can see that for leading order
$\eta-\eta_{B}=a_B^{-1}(t-t_B)$, where $\eta_B$ is the conformal
time at the bouncing point. Then from Eq. (\ref{qandcs2t}) one can
express $Q$ and $c_s^2$ w.r.t. $\eta$ approximately as:
\be\label{qandcs2eta}
Q\simeq-\frac{3t_0^2X_B}{\gamma^2\alpha}e^{-\chi
a_B^2(\eta-\eta_B)^2}~,~c_s^2\simeq-\frac{1}{2}\frac{1+2\alpha
a_B^2(\eta-\eta_B)^2}{1+3\alpha a_B^2(\eta-\eta_B)^2}~, \ee and
the parameter $z$ turns out to be: \be\label{zofeta}
z=a\frac{\sqrt{Q}}{c_s}\simeq
a_B\sqrt{\frac{6t_0^2X_B}{\gamma^2\alpha}}e^{\frac{a_B^2}{2}(\alpha-\chi)(\eta-\eta_B)^2}~.
\ee

Though $Q/c_s^2>0$ insures there is no ghost instability, $c_s^2<0$
leads to a gradient instability, which mainly affects the
small-scale perturbation modes. These small-scale modes are
within the horizon, i.e. the wavelength $\lambda<{\cal H}_{B-},{\cal
H}_{B+}$ (noting that the superhorizon modes are hardly affected,
see \cite{Koehn:2013upa}.), in which case the perturbation modes present
quantum-behavior. At classical level, it still remains an open issue of how to estimate the
effect of this gradient instability. Here, we will
not involve this issue in this paper, which is beyond our scope.

\begin{figure}[htbp]
\centering
\includegraphics[scale=0.6]{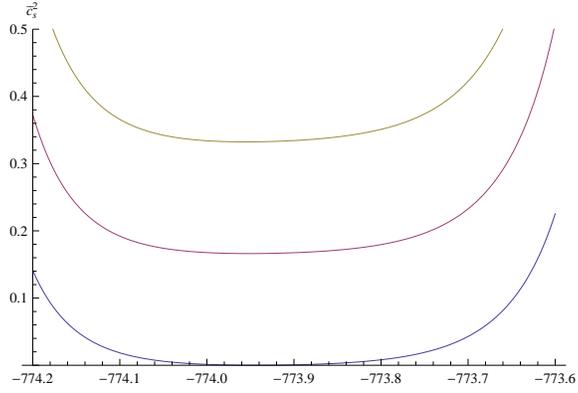}
\caption{Plot of the new sound speed squared under correction
(\ref{correction}) during bouncing phase. With the correction, the
sound speed squared can be kept above zero, avoiding the gradient
instability. Various examples are shown with $\xi_0$ set to be
$-1$ (blue), $-1.5$ (pink) and $-2$ (gold). At the edge of bounce
region where the correction term starts to decay, it will be
matched to values of $c_s^2$ in contracting/expanding region,
which is unity.}\label{cs2new}
\end{figure}

However, the nasty instability might be avoided if we expect some additional corrections in
the high energy region such as around the bounce
\cite{Pirtskhalava:2014esa}. For example, we phenomenologically assume that there is an additional term in the action of
gravity such as: \be\label{correction} \Delta
S^{(2)}=\frac{1}{2}\int d\eta d^3x a^2\xi(t)R_s^{(3)}~, \ee the physics of which is yet to be known, where
the 3-metric $R_s^{(3)}$ can be viewed as the scalar component of
potential term of the 4D gravity. In order not to affect gravity at other regions of evolution, $\xi(t)$ is
required to go close to zero at past or future limit but reach its
extreme value at the bounce. Since as a spatial curvature,
$R_s^{(3)}\sim (\partial\zeta)^2$ only, so this term will do
nothing to the background evolution as well as the time derivative
of $\zeta$, but will add another spatial derivative term of
$\zeta$ such as \be -\frac{1}{2}\int d\eta d^3x
a^2\left(Q\xi(t)+\frac{2M_p^2\dot\xi(t)}{M_p^2H-G_XX\dot\phi}\right)(\partial\zeta)^2~,
\ee so adding this term, the effective sound speed squared
becomes: \be
\bar{c}_{s}^2=c_s^2\left[1+\xi(t)+\frac{2M_p^2\dot\xi(t)}{Q(M_p^2H-G_XX\dot\phi)}\right]~.
\ee In order to make $\bar{c}_{s}^2>0$ around the bounce, we only
need \be
1+\xi(t)+\frac{2M_p^2\dot\xi(t)}{Q(M_p^2H-G_XX\dot\phi)}<0~, \ee
which can be easily satisfied by choosing proper form of $\xi(t)$.
As an example, we parameterize $\xi(t)$ as $\xi(t)=\xi_0
e^{-\lambda_\xi t^2}$, and plot the evolution of
$\bar{c}_{s}^2(t)$ in Fig. \ref{cs2new}. From the plot we can see
that now the sound speed squared will be positive all the time,
avoiding the gradient instability around the bounce, while reduce
to the one in contracting phase due to the quick decay of $\xi(t)$
far from the bounce.

The EoM of perturbation now becomes \be\label{eombouncing2}
u_k^{\prime\prime}+\left(\bar{c}_{s}^2k^2-(\alpha-\chi)a_B^2\right)u_k=0~,
\ee with a positive $\bar{c}_{s}^2$. As explicit examples, we set
$\xi_0=-1$, $-1.5$ and $-2$ while $\lambda_\xi=1$, and find that
$\bar{c}_s^2$ can be easily kept above zero (the minimum required
value of $\xi_0$ is $-1$, which gives $\bar{c}_s^2\simeq 0$).
%According to Eqs. (\ref{eompertscalar}), (\ref{qandcs2eta}) and (\ref{zofeta}), the equation of motion of the perturbations in bouncing phase reads (at the leading order):
%\be\label{eombouncing}
%u_k^{\prime\prime}-\left(\frac{1}{2}k^2+(\alpha-\chi)a_B^2\right)u_k=0~,
%\ee
%which gives
The solution is: \bea\label{solbouncing}
\label{solbouncingsub}%u_k&=&d_1e^{l(\eta-\eta_B)}+d_2e^{-l(\eta-\eta_B)}~,~k^2>2(\chi-\alpha)a_B^2~,\\
u_k&=&d_1\cos[l(\eta-\eta_B)]+d_2\sin[l(\eta-\eta_B)] \eea for any
$k$, where $l^2=\bar{c}_{s}^2k^2+(\alpha-\chi)a_B^2$. In general,
we have $\chi>\alpha$ during bouncing phase, and this is the
requirement of kinetic dominance in order to get a bounce:
$\ddot\phi>H\dot\phi$. This relation can also be numerically
checked in our model.

%Since $\bar{l}^2$ is now positive definite, there will be no
%exponential solutions of $u_k$ and $\zeta_k$.

%According to the mathematical properties of $M\left(0,-3/2,-\sqrt{2}k(\eta-\eta_B)\right)$ and $W\left(0,-3/2,-\sqrt{2}k(\eta-\eta_B)\right)$, one can get the solutions of subhorizon ($k|\eta|\gg1$) and superhorizon ($k|\eta|\ll1$) as:
%\bea\label{solbouncingsub}
%\zeta_k&\sim&\\
%\label{solbouncingsup}
%\zeta_k&\sim&
%\eea
%respectively. Note that for the subhorizon modes, the growth of the perturbation amplitude in the whole bouncing phase is:
%\be
%\zeta_k
%\ee
%while for the superhorizon modes \footnote{Since at the bouncing point the horizon is divergent, actually all the modes will enter into horizon at that time point. However for ultraviolet modes which stays outside horizon at the entrance/exit of the bouncing phase, the period of the subhorizon region can be neglected.} the growth of the perturbation amplitude in the whole bouncing phase is:
%\be
%\zeta_k
%\ee

The solution during bouncing phase should be matched onto the
solution in contracting phase via the matching conditions. We
apply the Hwang-Vishniac (Deruelle-Mukhanov) matching conditions
\cite{Hwang:1991an, Deruelle:1995kd} that $\zeta_k$ and its
derivative $\zeta_k^\prime$ should be matched at the pivot point
$\eta_{B-}$ (see also \cite{Nishi:2014bsa} for matching conditions
in Galileon theories.). Using the solutions
(\ref{solcontractingsub}), (\ref{c1c2}) and (\ref{solbouncingsub})
we have: \bea
d_1&=&\sqrt{\frac{2}{\pi\epsilon_{c}\bar{c}_{s}^{2}k}}\Bigg\{\sin[\bar{c}_{s}k(\eta_{B}-\eta_{B-})]\Big(c_{2}\cos[k(\eta_{B-}-\tilde{\eta}_{B-})]-c_{1}\sin[k(\eta_{B-}-\tilde{\eta}_{B-})]\Big)\nonumber\\ &&+\bar{c}_{s}\cos[\bar{c}_{s}k(\eta_{B}-\eta_{B-})]\Big(c_{1}\cos[k(\eta_{B-}-\tilde{\eta}_{B-})]+c_{2}\sin[k(\eta_{B-}-\tilde{\eta}_{B-})]\Big)\Bigg\}~,\nonumber\\
d_2&=&\sqrt{\frac{2}{\pi\epsilon_{c}\bar{c}_{s}^{2}k}}\Bigg\{\cos[\bar{c}_{s}k(\eta_{B}-\eta_{B-})]\Big(c_{2}\cos[k(\eta_{B-}-\tilde{\eta}_{B-})]-c_{1}\sin[k(\eta_{B-}-\tilde{\eta}_{B-})]\Big)\nonumber\\
&&-\bar{c}_{s}\sin[\bar{c}_{s}k(\eta_{B}-\eta_{B-})]\Big(c_{1}\cos[k(\eta_{B-}-\tilde{\eta}_{B-})]+c_{2}\sin[k(\eta_{B-}-\tilde{\eta}_{B-})]\Big)\Bigg\}
\label{d1d2sub}\eea for subhorizon modes while with
(\ref{solcontractingsup}), (\ref{c1c2}) and (\ref{solbouncingsub})
we have: \bea\label{d1d2sup}
d_1&=&\frac{[(\epsilon_c-1){\cal H}_{B-}]^\frac{1}{1-\epsilon_c}}{2^{\nu_-}\Gamma(1+\nu_-)\sqrt{\epsilon_c}}k^{-\frac{1}{2}+\frac{1}{-1+\epsilon_c}}\cos[l(-\eta_B+\eta_{B-})]c_1~,\nonumber\\
d_2&=&\frac{[(\epsilon_c-1){\cal H}_{B-}]^\frac{1}{1-\epsilon_c}}{2^{\nu_-}\Gamma(1+\nu_-)\sqrt{\epsilon_c}}k^{-\frac{1}{2}+\frac{1}{-1+\epsilon_c}}\sin[l(-\eta_B+\eta_{B-})]c_1~
\eea
for superhorizon modes.
\subsubsection{expanding phase}
After the bounce, the universe will enter into an inflationary expanding phase, with the Hubble parameter being a little bit decreasing but nearly steady. In this phase, all the features of the shape functions will decay again and make them the same constant value as before the bounce. Moreover, the parameters $Q$ and $c_s^2$ will become constant again, with $Q\simeq2M_p^2\epsilon_e$ and $c_s^2\simeq 1$. The scale factor during this time can be parameterized as:
\be
a(\eta)\sim (\eta-\tilde\eta_{B+})^\frac{1}{\epsilon_e-1}~,
\ee
while $\tilde\eta_{B+}\equiv\eta_{B+}-[(\epsilon_e-1){\cal H}]^{-1}$ to protect the continuity of ${\cal H}$, and the slow-roll parameter $\epsilon_e$ is defined in Eq. (\ref{epsilon}). However, since during the most of the time $\phi$ is much smaller than the expectation value $v$, one have $\epsilon_e\ll 1$ and $\dot\epsilon_e\ll H\epsilon_e$, thus it is reasonable to take $\epsilon_e$ as nearly a constant. Then the equation of motion (\ref{eompertscalar}) becomes:
\be
u_k^{\prime\prime}+k^2u-\frac{2-\epsilon_e}{(1-\epsilon_e)^2}\frac{u}{(\eta-\tilde\eta_{B+})^2}=0~,
\ee
and the solution is the Hankal function:
\be\label{solexpanding}
u_k=\sqrt{|\eta-\tilde\eta_{B+}|}[g_1(k)J_{\nu_+}(k|\eta-\tilde\eta_{B+}|)+g_2(k)J_{-\nu_+}(k|\eta-\tilde\eta_{B+}|)]~,
\ee
where
\be
\nu_+=\frac{3-\epsilon_e}{2(\epsilon_e-1)}~.
\ee

As has been discussed in the subsection 2, the solution of Eq. (\ref{solexpanding}) also has two branches of solution. One is the subhorizon solution:
\bea\label{solexpandingsub}
u_k&\simeq&g_1(k)\sqrt{\frac{2}{\pi k}}\cos(k|\eta-\tilde\eta_{B+}|)~,~g_2(k)\sqrt{\frac{2}{\pi k}}\sin(k|\eta-\tilde\eta_{B+}|)~,\nonumber\\
\zeta_k&\simeq&\frac{g_1(k)}{a_{B+}}\sqrt{\frac{2}{\pi\epsilon_e k}}\left(\frac{\eta-\tilde\eta_{B+}}{\eta_{B+}-\tilde\eta_{B+}}\right)^{\frac{1}{1-\epsilon_e}}\cos(k|\eta-\tilde\eta_{B+}|)~,~\frac{g_2(k)}{a_{B+}}\sqrt{\frac{2}{\pi\epsilon_e k}}\left(\frac{\eta-\tilde\eta_{B+}}{\eta_{B+}-\tilde\eta_{B+}}\right)^{\frac{1}{1-\epsilon_e}}\sin(k|\eta-\tilde\eta_{B+}|)~.
\eea
This solution is for the large $k$ modes, which still remains inside the horizon at least at the time when the universe enters the expanding phase. This branch should be matched with the large $k$ modes in bouncing phase (\ref{solbouncingsub}). The other branch of the solution of (\ref{solexpanding}) is the superhorizon solution:
\bea\label{solexpandingsup}
u_k&\sim&\frac{g_1(k)}{2^{\nu_+}\Gamma(\nu_++1)\sqrt{k}}(k|\eta-\tilde\eta_{B+}|)^\frac{1}{\epsilon_e-1}~,~\frac{g_2(k)}{2^{-\nu_+}\Gamma(-\nu_++1)\sqrt{k}}(k|\eta-\tilde\eta_{B+}|)^\frac{\epsilon_e-2}{\epsilon_e-1}~,\nonumber\\
\zeta_k&\sim& \frac{g_1(k)[(\epsilon_e-1){\cal
H}_{B+}]^\frac{1}{1-\epsilon_e}}{2^{\nu_+}\Gamma(\nu_++1)\sqrt{\epsilon_e}a_{B+}}k^{-\frac{\epsilon_e-3}{2(\epsilon_e-1)}}~,~\frac{g_2(k)[(\epsilon_e-1){\cal
H}_{B+}]^\frac{1}{1-\epsilon_e}}{2^{-\nu_+}\Gamma(-\nu_++1)\sqrt{\epsilon_e}a_{B+}}k^{-\frac{\epsilon_e-3}{2(\epsilon_e-1)}}(k|\eta-\tilde\eta_{B+}|)^\frac{\epsilon_e-3}{\epsilon_e-1}~,
\eea This solution has two origins: one of the origin is from the
large $k$ modes, which exit the horizon during the expanding
phase. Since it shares the same equation with the subhorizon
solution (\ref{solexpandingsub}), the coefficients of the two
parts will be the same. The other origin is from the small $k$
modes, which already exit the horizon when the universe enters the
expanding phase. This solution should be matched with the small
$k$ modes in bouncing phase (\ref{d1d2sub}). So we have: \bea
g_1&=&\frac{1}{2\bar{c}_{s}}\sqrt{\frac{\epsilon_{e}}{\epsilon_{c}}}\Bigg\{2\bar{c}_{s}\cos(\bar{c}_{s}k\Delta\eta_{B})\Big(c_{1}\cos[k(\eta_{B-}-\tilde{\eta}_{B-}-\eta_{B+}+\tilde{\eta}_{B+})]+c_{2}\sin[k(\eta_{B-}-\tilde{\eta}_{B-}-\eta_{B+}+\tilde{\eta}_{B+})]\Big)\nonumber\\ &&-(\bar{c}_{s}^{2}+1)\sin(\bar{c}_{s}k\Delta\eta_{B})\Big(c_{2}\cos[k(\eta_{B-}-\tilde{\eta}_{B-}-\eta_{B+}+\tilde{\eta}_{B+})]-c_{1}\sin[k(\eta_{B-}-\tilde{\eta}_{B-}-\eta_{B+}+\tilde{\eta}_{B+})]\Big)\nonumber\\ &&+(\bar{c}_{s}^{2}-1)\sin(\bar{c}_{s}k\Delta\eta_{B})\Big(c_{2}\cos[k(\eta_{B-}-\tilde{\eta}_{B-}+\eta_{B+}-\tilde{\eta}_{B+})]-c_{1}\sin[k(\eta_{B-}-\tilde{\eta}_{B-}+\eta_{B+}-\tilde{\eta}_{B+})]\Big)\Bigg\}~,\nonumber\\
g_2&=&\frac{1}{2\bar{c}_{s}}\sqrt{\frac{\epsilon_{e}}{\epsilon_{c}}}\Bigg\{2\bar{c}_{s}\cos(\bar{c}_{s}k\Delta\eta_{B})\Big(c_{2}\cos[k(\eta_{B-}-\eta_{B+}-\tilde{\eta}_{B-}+\tilde{\eta}_{B+})]-c_{1}\sin[k(\eta_{B-}-\eta_{B+}-\tilde{\eta}_{B-}+\tilde{\eta}_{B+})]\Big)\nonumber\\
&&+(\bar{c}_{s}^{2}+1)\sin(\bar{c}_{s}k\Delta\eta_{B})\Big(c_{1}\cos[k(\eta_{B-}-\eta_{B+}-\tilde{\eta}_{B-}+\tilde{\eta}_{B+})]+c_{2}\sin[k(\eta_{B-}-\eta_{B+}-\tilde{\eta}_{B-}+\tilde{\eta}_{B+})]\Big)\nonumber\\
&&+(\bar{c}_{s}^{2}-1)\sin(\bar{c}_{s}k\Delta\eta_{B})\Big(c_{1}\cos[k(\eta_{B-}+\eta_{B+}-\tilde{\eta}_{B-}-\tilde{\eta}_{B+})]+c_{2}\sin[k(\eta_{B-}+\eta_{B+}-\tilde{\eta}_{B-}-\tilde{\eta}_{B+})]\Big)\Bigg\}~
\eea for those which exit the horizon in expanding phase, where we define $\Delta\eta_B\equiv\eta_{B+}-\eta_{B-}$, and
\bea\label{g1g2sup}
g_1&=&\frac{2^{\nu_+-\nu_-}\Gamma(1+\nu_+)}{(\epsilon_e-3)\Gamma(1+\nu_-)}\sqrt{\frac{\epsilon_e}{\epsilon_c}}\frac{[(\epsilon_c-1){\cal H}_{B-}]^\frac{1}{1-\epsilon_c}}{[(\epsilon_e-1){\cal H}_{B+}]^\frac{1}{1-\epsilon_e}}[(\epsilon_e-3)\cos(l\Delta\eta_{B})+l{\cal H}_{B+}\sin(l\Delta\eta_{B})]c_1 k^{\frac{1}{\epsilon_c-1}+\frac{1}{1-\epsilon_e}}~,\nonumber\\
g_2&=&\frac{2^{-\nu_+-\nu_-}\Gamma(1-\nu_+)}{(\epsilon_e-3)\Gamma(1+\nu_-)}\sqrt{\frac{\epsilon_e}{\epsilon_c}}[(\epsilon_c-1){\cal H}_{B-}]^\frac{1}{1-\epsilon_c}[(\epsilon_e-1){\cal H}_{B+}]^\frac{1}{1-\epsilon_e}l(\epsilon_e-1)\sin(l\Delta\eta_{B})c_1 k^{-1+\frac{1}{\epsilon_c-1}-\frac{1}{1-\epsilon_e}}
\eea
for those which exit the horizon before the bounce.

We are observing superhorizon modes, namely. For the solution of
the superhorizon mode (\ref{solexpandingsup}) which we're
observing, the constant branch $g_1$ will be dominant over the
decaying branch $g_2$. The power spectrum is then \be
P_\zeta\equiv\frac{k^3}{2\pi^2}|\zeta|^2=\frac{k^3}{2\pi^2}|\frac{g_1(k)[(\epsilon_e-1){\cal
H}_{B+}]^\frac{1}{1-\epsilon_e}}{2^{\nu_+}\Gamma(\nu_++1)\sqrt{2\epsilon_e}a_{B+}}k^{-\frac{\epsilon_e-3}{2(\epsilon_e-1)}}|^2~.
\ee

Considering that $c_1=\sqrt{\pi}/2$, $c_1=i\sqrt{\pi}/2$, the
modulus squared of $g_1(k)$ is calculated as: \bea
|g_{1}(k)|^{2}&=&\frac{\pi}{16\bar{c}_{s}^{2}}\frac{\epsilon_{e}}{\epsilon_{c}}\Bigg\{(1+\bar{c}_{s}^{2})^{2}+(1-\bar{c}_{s}^{4})\cos\left(\frac{2k}{(\epsilon_{e}-1)\mathcal{H}_{B+}}\right)-(1-\bar{c}_{s}^{2})^2\cos(2\bar{c}_{s}k\Delta\eta_{B})\nonumber\\
&&-(1-\bar{c}_{s}^{2})\left[(1+\bar{c}_{s}^{2})\cos\left(\frac{2k}{(\epsilon_{e}-1)\mathcal{H}_{B+}}\right)\cos(2\bar{c}_{s}k\Delta\eta_{B})+2\bar{c}_{s}\sin\left(\frac{2k}{(\epsilon_{e}-1)\mathcal{H}_{B+}}\right)\sin(2\bar{c}_{s}k\Delta\eta_{B})\right]\Bigg\}
\eea For large $k$ modes, the trigonometric functions in the above
expression will oscillate rapidly, showing only their average
values, which is zero. Then the mean value of
$|g_{1}(k)|^{2}\simeq(\pi/16\bar{c}_{s}^{2})(\epsilon_{e}/\epsilon_{c})(1+\bar{c}_{s}^{2})^{2}$.

One can see that, for modes that exit horizon before bounce,
\bea\label{spectrumscalarsup}
P^{con}_\zeta&=&\frac{(\epsilon_{c}-1)^{\frac{2}{1-\epsilon_{c}}}H_{B-}^{2}[(\epsilon_{e}-3)\cos(l\Delta\eta_{B})+l{\cal
H}_{B+}^{-1}\sin(l\Delta\eta_{B})]^{2}}{2^{2\nu_{-}+4}\pi(\epsilon_{e}-3)^{2}\Gamma^{2}(1+\nu_{-})\epsilon_{c}}\left(\frac{k}{{\cal
H}_{B-}}\right)^{\frac{2\epsilon_{c}}{\epsilon_{c}-1}} \eea with
the index \be\label{indexscalarsup}
n^{con}_\zeta\simeq\frac{2\epsilon_c}{\epsilon_{c}-1}~. \ee Since
$\epsilon_c\geq3$, it will have a blue tilt.

While for modes that exit horizon after bounce, the power spectrum
$P_\zeta(k)$ is in the form of: \be\label{spectrumscalarsub}
P^{inf}_\zeta\simeq\frac{(1+\bar{c}_{s}^{2})^{2}(\epsilon_{e}-1)^{\frac{2}{1-\epsilon_{e}}}H_{B+}^2}{64\pi\epsilon_{c}\bar{c}_{s}^{2}2^{2\nu_{+}}\Gamma^{2}(\nu_{+}+1)}\left(\frac{k}{{\cal
H}_{B+}}\right)^{\frac{2\epsilon_{e}}{\epsilon_{e}-1}} \ee and the
spectral index \be\label{indexscalarsub}
n^{inf}_\zeta=1+\frac{2\epsilon_e}{\epsilon_e-1}~. \ee We can see
that, taking the correction (\ref{correction}), it shows
scale-invariance even at the $k\rightarrow \infty$ limit, which is
consistent with the data. When we take the slow-roll limit
$\epsilon_e\simeq0$, we have
$P_\zeta\simeq(1+\bar{c}_{s}^{2})^{2}H_{B+}^2/(32\pi^2\epsilon_{c}\bar{c}_{s}^{2})$.
Moreover, if one also take into account the corrections from the
trigonometric functions in $|g_{1}(k)|^{2}$, both $P_\zeta$ and
$n_\zeta$ will have some wiggling behavior due to such a
correction, which will be observed if we have enough accurate
data. The same property has been discussed in the second paper of
Ref. \cite{Cai:2007zv}. This property can be used to distinguish
from the standard inflation models without preceding evolution.

%\subsubsection{corrections from higher energy regions}

%Finally, since in inflationary expanding phase a conserved quantity at large scales, what we observe at the reentering of the horizon is just that exit (or remain outside) the horizon. From this we can get the amplitude of the scalar perturbations observed by us is:
%\bea
%P^{con}_\zeta&=&\frac{H_{B-}^2}{8\pi^2M_p^2\epsilon}\left(\frac{k}{k_{B-}}\right)^{n^{con}_s-1}~~~~~\text{for small } k~,\\
%P^{inf}_\zeta&=&\frac{H^2}{8\pi^2M_p^2\epsilon}~~~~~\text{for large } k~,
%\eea
%and the spectral index is:
%\bea
%n_s^{con}&=&1+\frac{2\epsilon}{\epsilon-1}~~~~~\text{for small } k~, \\
%n_s^{inf}&=&1+\frac{2\epsilon}{\epsilon-1}\simeq1-2\epsilon~~~~~\text{for large } k~,
%\eea

\subsection{tensor perturbations}
Besides the scalar part, there are also tensor part perturbations. Since the field is pure scalar field and there are no tensor degrees of freedom in the field part, the tensor part totally come from the perturbations of the gravity. It is believed that the tensor perturbation is originated from the primordial gravitational waves, and although very weak, the recent gravitational wave explorers such as BICEP has shed some light on our future ability of detecting them more clearly \cite{Ade:2014xna}. As a theoretical study, here we also discuss the tensor perturbations generated in our model. According to \cite{Mukhanov:1990me}, the equation of motion for tensor perturbations can be written as:
\be\label{pertlagrangian2}
{\cal S}_T^{(2)}=\frac{1}{2}\int d\eta d^3xa^2\Bigl[h^{\prime 2}-(\partial h)^2\Bigr]~,
\ee
where $h$ is the amplitude of the spatial-spatial part perturbations of the metric, namely $h_{ij}$. There are two degrees of freedom in tensor parts, namely $h_+$ and $h_\times$, obeying the same action and equation of motion. From (\ref{pertlagrangian2}) we get the equation of motion for $h$ as:
\be\label{eomperttensor}
v_k^{\prime\prime}+(k^2-\frac{a^{\prime\prime}}{a})v_k=0~,
\ee
where we define $v\equiv ah/2$. From both the action and the equation of motion we can see that, the evolution of the tensor part will be affected only by the evolution of the universe itself, namely the scale factor $a$, rather than the functions containing $\phi$. This is due to that tensor part is decoupled from scalar part, and only comes from the gravity. Thus solving the equation of motion for tensor perturbations will be much easier than that for scalar ones.

We follow the same procedure as what we did in subsec. {\bf A}. At the very beginning when all the modes are deep inside the horizon, the last term in the left hand side of Eq. (\ref{eomperttensor}) is not important, thus similar to scalar part, $v$ has the oscillating solution. For the initial condition of $h$, we also choose the Bunch-Davies vacuum, so the tensor perturbation at the initial time have the solution as:
\be\label{ict}
v_k=\frac{1}{\sqrt{2k}}e^{i k\eta}~.
\ee

For contracting phase where we have $a(\eta)\sim(\eta-\tilde\eta_{B-})^{\frac{\epsilon_c}{\epsilon_c-1}}$, we have the solution for tensor perturbations as:
\be\label{solcontractingten}
v_k=\sqrt{|\eta-\tilde\eta_{B-}|}[c^T_1(k)J_{\nu_-}(k|\eta-\tilde\eta_{B-}|)+c^T_2(k)J_{-\nu_-}(k|\eta-\tilde\eta_{B-}|)]~,
\ee
where $\nu_-$ is already given in Eq. (\ref{numinus}). Since in this period the functions such as $Q$ and $c_s^2$ for scalar perturbations also becomes trivial, the solution of tensor and scalar perturbations are actually very similar. For large $k$ modes which remains inside horizon during contraction phase, the solution is:
\bea
v_k&\simeq&c^T_{1}(k)\sqrt{\frac{2}{\pi k}}\cos(k|\eta-\tilde\eta_{B-}|)~,~c^T_{2}(k)\sqrt{\frac{2}{\pi k}}\sin(k|\eta-\tilde\eta_{B-}|)~, \nonumber\\
h_k&\simeq&\frac{c^T_{1}(k)}{a_{B-}}\sqrt{\frac{8}{\pi k}}\left(\frac{\eta-\tilde\eta_{B-}}{\eta_{B-}-\tilde\eta_{B-}}\right)^{\frac{1}{1-\epsilon_c}}\cos(k|\eta-\tilde\eta_{B-}|)~,~\frac{c^T_{2}(k)}{a_{B-}}\sqrt{\frac{8}{\pi k}}\left(\frac{\eta-\tilde\eta_{B-}}{\eta_{B-}-\tilde\eta_{B-}}\right)^{\frac{1}{1-\epsilon_c}}\sin(k|\eta-\tilde\eta_{B-}|)~,
\eea
while for small $k$ modes which has already exit the horizon during contracting phase, the solution will be
\bea
v_k&\simeq&\frac{c^T_1(k)}{2^{\nu_-}\Gamma(\nu_-+1)\sqrt{k}}(k|\eta-\tilde\eta_{B-}|)^\frac{1}{\epsilon_c-1}~,~\frac{c^T_2(k)}{2^{-\nu_-}\Gamma(-\nu_-+1)\sqrt{k}}(k|\eta-\tilde\eta_{B-}|)^\frac{\epsilon_c-2}{\epsilon_c-1}~,\nonumber\\
h_k&\simeq& \frac{c^T_1(k)[(\epsilon_c-1){\cal H}_{B-}]^{\frac{1}{1-\epsilon_c}}}{2^{\nu_--1}\Gamma(\nu_-+1)a_{B-}}k^{-\frac{\epsilon_c-3}{2(\epsilon_c-1)}}~,~\frac{c^T_2(k)[(\epsilon_c-1){\cal H}_{B-}]^{\frac{1}{1-\epsilon_c}}}{2^{-\nu_--1}\Gamma(-\nu_-+1)a_{B-}}k^{-\frac{\epsilon_c-3}{2(\epsilon_c-1)}}(k|\eta-\tilde\eta_{B-}|)^\frac{\epsilon_c-3}{\epsilon_c-1}~.
\eea
By connecting the subhorizon solution with initial condition (\ref{ict}), one can get the coefficients:
\be
c^T_1=\frac{\sqrt{\pi}}{2}~,~c^T_2=i\frac{\sqrt{\pi}}{2}~.
\ee

For bouncing phase however, the tensor perturbation shows greatest difference with the scalar perturbations, for it is independent of functions such as $Q$ and $c_s^2$, which becomes nontrivial any more in bouncing phase. Considering Eq. (\ref{eomperttensor}) and the expression of scale factor in bouncing phase (\ref{abounce}), the equation of motion for $v_k$ now becomes (in its leading order):
\be
v_k^{\prime\prime}+(k^2-\alpha a_B^2)v_k=0~,
\ee
and the solution is:
\bea
\label{solbouncingsubten}v_k&=&d^T_1a_B^{-1}\cos[l(\eta-\eta_B)]+d^T_2a_B^{-1}\sin[l(\eta-\eta_B)]~,~k>\sqrt{\alpha}a_B~,\\
\label{solbouncingsupten}v_k&=&d^T_1e^{l(\eta-\eta_B)}+d^T_2e^{-l(\eta-\eta_B)}~,~k<\sqrt{\alpha}a_B~,
\eea
where $l^2\equiv|k^2-\alpha a_B^2|$. Note that $\eta$ will run from $\eta_{B-}$ to $\eta_{B-}$, and the value of $v_k$ at $\eta=\eta_{B-}$ is the amplitude growth of the tensor perturbations during bouncing period. Via matching conditions which requires continuity of $h_k$ and $h_k^\prime$ \cite{Hwang:1991an, Deruelle:1995kd}, one can determine the coefficient as:
\bea
d^T_1&\simeq&\sqrt{\frac{2}{\pi k}}[c_1^T\cos(k(\eta_B-\tilde\eta_{B-}))+c_2^T\sin(k(\eta_B-\tilde\eta_{B-}))]~,\nonumber\\
d^T_2&\simeq&\sqrt{\frac{2}{\pi k}}[c_2^T\cos(k(\eta_B-\tilde\eta_{B-}))-c_1^T\sin(k(\eta_B-\tilde\eta_{B-}))]~
\eea
for subhorizon modes, while
\bea
d^T_1&\simeq&\frac{c_1^T[(\epsilon_c-1){\cal H}_{B-}]^{\frac{1}{1-\epsilon_c}}}{2^{1+\nu_-}l\Gamma(1+\nu_-)}e^{l(\eta_B-\eta_{B-})}k^{-\frac{\epsilon_c-3}{2(\epsilon_c-1)}}~,\nonumber\\
d^T_2&\simeq&\frac{c_1^T[(\epsilon_c-1){\cal H}_{B-}]^{\frac{1}{1-\epsilon_c}}}{2^{1+\nu_-}l\Gamma(1+\nu_-)}e^{-l(\eta_B-\eta_{B-})}k^{-\frac{\epsilon_c-3}{2(\epsilon_c-1)}}~
\eea
for superhorizon modes.

In expanding phase, the tensor perturbations becomes concordance with the scalar ones again. Since the scale factor now becomes $a(\eta)\sim (\eta-\tilde\eta_{B+})^{1/(1-\epsilon_e)}$, the solution for $v_k$ becomes:
\bea
v_k=\sqrt{|(\eta-\tilde\eta_{B+})|}[g^T_1(k)J_{\nu_+}(k|\eta-\tilde\eta_{B+}|)+g^T_2(k)J_{-\nu_+}(k|\eta-\tilde\eta_{B+}|)]~,
\eea
Also the solutions are divided by subhorizon and superhorizon parts. The subhorizon part is:
\bea\label{solexpandingsubten}
v_k&\simeq&g^T_{1}(k)\sqrt{\frac{2}{\pi k}}\cos(k|\eta-\tilde\eta_{B+}|)~,~g^T_{2}(k)\sqrt{\frac{2}{\pi k}}\sin(k|\eta-\tilde\eta_{B+}|)~,\nonumber\\
h_k&\simeq&\frac{g^T_{1}(k)}{a_{B+}}\sqrt{\frac{8}{\pi k}}\left(\frac{\eta-\tilde\eta_{B+}}{\eta_{B+}-\tilde\eta_{B+}}\right)^{\frac{1}{1-\epsilon_e}}\cos(k|\eta-\tilde\eta_{B+}|)~,~\frac{g^T_{2}(k)}{a_{B+}}\sqrt{\frac{8}{\pi k}}\left(\frac{\eta-\tilde\eta_{B+}}{\eta_{B+}-\tilde\eta_{B+}}\right)^{\frac{1}{1-\epsilon_e}}\sin(k|\eta-\tilde\eta_{B+}|)~.
\eea
which denotes the modes that remains inside the horizon at the beginning of the expanding phase. While the superhorizon parts is:
\bea\label{solexpandingsupten}
v_k&\simeq&\frac{g^T_1(k)}{2^{\nu_+}\Gamma(\nu_++1)\sqrt{k}}(k|\eta-\tilde\eta_{B+}|)^\frac{1}{\epsilon_e-1}~,~\frac{g^T_2(k)}{2^{-\nu_+}\Gamma(-\nu_++1)\sqrt{k}}(k|\eta-\tilde\eta_{B+}|)^\frac{\epsilon_e-2}{\epsilon_e-1}~,\nonumber\\
h_k&\simeq&\frac{g^T_1(k)[(\epsilon_e-1){\cal H}_{B+}]^{\frac{1}{1-\epsilon_e}}}{2^{\nu_+-1}\Gamma(\nu_++1)a_{B+}}k^{-\frac{\epsilon_e-3}{2(\epsilon_e-1)}}~,~\frac{g^T_2(k)[(\epsilon_e-1){\cal H}_{B+}]^{\frac{1}{1-\epsilon_e}}}{2^{-\nu_+-1}\Gamma(-\nu_++1)a_{B+}}k^{-\frac{\epsilon_e-3}{2(\epsilon_e-1)}}(k|\eta-\tilde\eta_{B+}|)^\frac{\epsilon_e-3}{\epsilon_e-1}~,
\eea
which denotes the modes that are outside the horizon. These modes also have two origins. For those which exit the horizon in expanding phase, using matching conditions with the solution in (\ref{solbouncingsubten}) and (\ref{solexpandingsubten}), we have:
\bea
g_1^T&\simeq&c_1\cos(k(\tilde{\eta}_{B+}-\tilde\eta_{B-}))+c_2\sin(k(\tilde{\eta}_{B+}-\tilde\eta_{B-}))~,\nonumber\\
g_2^T&\simeq&c_2\cos(k(\tilde{\eta}_{B+}-\tilde\eta_{B-}))-c_1\sin(k(\tilde{\eta}_{B+}-\tilde\eta_{B-}))~,
\eea
and for those which exit the horizon before the bounce, one obtain
\bea
g_1^T&\simeq&\frac{2^{\nu_+-\nu_-}\Gamma(1+\nu_+)}{(\epsilon_e-3)\Gamma(1+\nu_-)}\frac{[(\epsilon_c-1){\cal H}_{B-}]^{\frac{1}{1-\epsilon_c}}}{[(\epsilon_e-1){\cal H}_{B+}]^{\frac{1}{1-\epsilon_e}}}\left((\epsilon_e-3)\cosh(l\Delta\eta_{B})+\frac{l}{{\cal H}_{B+}}\sinh(l\Delta\eta_{B})\right)c_1 k^{\frac{1}{\epsilon_c-1}+\frac{1}{1-\epsilon_e}}~,\nonumber\\
g_2^T&\simeq&\frac{2^{-\nu_+-\nu_-}\Gamma(1-\nu_+)}{(\epsilon_e-3)\Gamma(1+\nu_-)}[(\epsilon_c-1){\cal H}_{B-}]^{\frac{1}{1-\epsilon_c}}[(\epsilon_e-1){\cal H}_{B+}]^{\frac{1}{1-\epsilon_e}}l(\epsilon_e-1)\sinh(l\Delta\eta_{B})c_1 k^{-1+\frac{1}{\epsilon_c-1}-\frac{1}{1-\epsilon_e}}~.
\eea

Similar as the scalar part, we are observing superhorizon modes, namely (\ref{solexpandingsupten}). For this solution, it has two branches, one is constant ($g_1$) and the other is decaying $g_2$, therefore the first one is dominant. The power spectrum is then
\be
P_T\equiv2\frac{k^3}{2\pi^2}|h_k|^2=\frac{k^3}{\pi^2}|\frac{g^T_1(k)[(\epsilon_e-1){\cal H}_{B+}]^{\frac{1}{1-\epsilon_e}}}{2^{\nu_+}\Gamma(\nu_++1)a_{B+}}k^{-\frac{\epsilon_e-3}{2(\epsilon_e-1)}}|^2~,
\ee
where the factor $2$ appears because there are 2 independent polarizations of the graviton \cite{Bassett:2005xm}. One can see that, for modes that exit horizon in the inflation phase,
\be
P_T=\frac{(\epsilon_{e}-1)^{\frac{2}{1-\epsilon_{e}}}H_{B+}^2}{2^{2\nu_{+}}\pi\Gamma^{2}(\nu_{+}+1)}\left(\frac{k}{{\cal H}_{B+}}\right)^{\frac{2\epsilon_{e}}{\epsilon_{e}-1}}~
\ee
with the index $n_T\simeq2\epsilon_e/(\epsilon_e-1)$. For $\epsilon_e\simeq 0$, we have $P_T=2(H_{B+}/\pi)^2$ and $n_T\simeq0$. While for modes that exit horizon before bounce,
\bea
P_T&=&[(\epsilon_{c}-1)]^{\frac{2}{1-\epsilon_{c}}}H_{B-}^2\frac{[(\epsilon_{e}-3)\cosh(l\Delta\eta_{B})+l{\cal H}_{B+}^{-1}\sinh(l\Delta\eta_{B})]^{2}}{2^{2\nu_{-}}\pi(\epsilon_{e}-3)^{2}\Gamma^{2}(1+\nu_{-})}\left(\frac{k}{{\cal H}_{B-}}\right)^{\frac{2\epsilon_{c}}{\epsilon_{c}-1}}
\eea
with the index $n_T\simeq2\epsilon_c/(\epsilon_{c}-1)$. Since $\epsilon_c\geq3$, it will have a blue tilt.

From the above result for scalar and tensor perturbations, we can also calculate the tensor-to-scalar ratio $r$, defined as $r\equiv P_T/P_\zeta$. The 1st year Planck paper suggest to have $r<0.1$ at $2\sigma$ level, while BICEP2 data shed some light on the possibility to measure it more accurately. In our model, we have
\be
r=\frac{64\epsilon_{c}\bar{c}_{s}^{2}}{(1+\bar{c}_{s}^{2})^{2}}~~~\text{for large}~k~,~~~r=16\epsilon_c~~~\text{for small}~k~.
\ee
One can see that for large $k$ modes which is observable, $r$ carries information from bouncing or contracting phase, which makes bounce detectable. For example, the Planck constraint $r<0.1$ requires that $\bar{c}_s^2\lesssim5.2\times10^{-4}$ for the minimum value of $\epsilon_c=3$, which furtherly lead to $0>(1+\xi_0)>-3\times10^{-4}$ for $\xi_0$ in the parametrization of $\xi(t)$. However, for small $k$ modes which has not been detectable yet, $r$ could be very large $(\gtrsim 50)$. This could be a smoking gun to the bounce inflation scenario.

%Finally, we can get the amplitude of the power spectrum for tensor perturbations as:
%\bea
%P_T&=&\frac{2H_{B-}^2}{\pi^2M_p^2}\left(\frac{k}{k_{B-}}\right)^{n_T}~~~~~\text{for small } k~,\\
%P_T&=&\frac{2H^2}{\pi^2M_p^2}~~~~~\text{for large } k~,
%\eea
%and its index is:
%\bea
%2\leq n^{con}_T&=&\frac{2\epsilon}{\epsilon-1}\leq 3~~~~~\text{for small } k~,\\
%n^{inf}_T&=&\frac{2\epsilon}{\epsilon-1}\simeq-2\epsilon~~~~~\text{for large } k~,
%\eea
As a side remark, we would like to comment that actually all the above analysis is based on the consideration that the bounce is not far from observable inflation, i.e. the e-fold number is not too much, so the information from the bounce can be detectable. However, there is also possibility that if the inflation was too long, therefore the observed modes comes from Bunch-Davies vacuum even in expanding time (like the purple line in the Fig. \ref{horizon}). In this case, we instead have $g_1=\sqrt{\pi}/2$, $g_2=i\sqrt{\pi}/2$, so everything could be completely the same as pure inflation and bounce becomes non-detectable. As for in the region before the bounce, this mode should be inside quantum phase all the time, and solutions in different stages may be matched quantum mechanically. Of course such an issue goes far beyond the scope of this paper, and actually we don't know how to do it yet. There may be chances that we pick up this topic in future investigations.

\section{fitting the data}
\begin{figure}[htbp]
\centering
\includegraphics[scale=0.6]{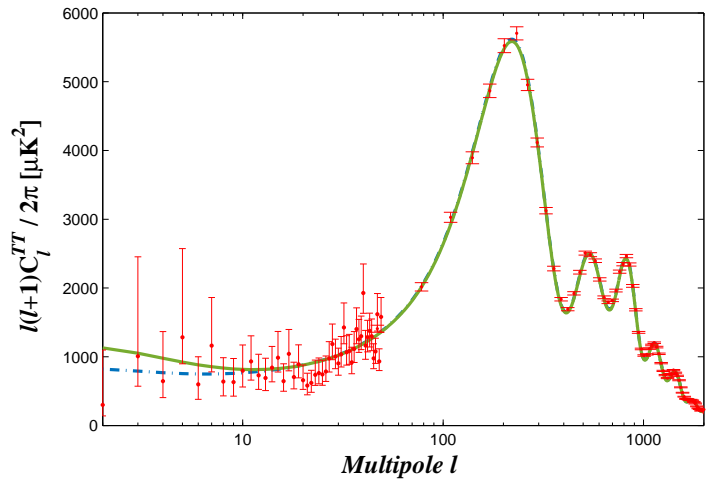}
\caption{Best-fit TT power spectra for pure power law (green solid) and our model (dashed) using Planck+WP data. The parameter choices are: $\Omega_bh^2=0.02203$, $\Omega_ch^2=0.1204$, $\tau=0.09$, $100\Theta_s=1.04$, $A_s=2.1\times10^{-9}$, $n_s=0.961$, $\epsilon_e=0.02$. The red points show the Planck data with 1$\sigma$ errors.}\label{TT}
\end{figure}

In this section, we try to fit our model with the newest released PLANCK data. Following (\ref{spectrumscalarsup}) and (\ref{spectrumscalarsub}), we parametrize the whole power spectrum as:
\bea
P_\zeta^{para}&=&\left[1-\tanh\left(\frac{k}{{\cal H}_{B-}}-1\right)\right]{\cal A}^{con}_\zeta\left(\frac{k}{{\cal H}_{B-}}\right)^{n^{con}_\zeta-1}+\left[1+\tanh\left(\frac{k}{{\cal H}_{B-}}-1\right)\right]{\cal A}^{inf}_\zeta\left(\frac{k}{{\cal H}_{B+}}\right)^{n^{inf}_\zeta-1}~,
\eea
where ${\cal A}^{con}_\zeta$ and ${\cal A}^{inf}_\zeta$ are the $k$-independent prefactors of expressions (\ref{spectrumscalarsup}) and (\ref{spectrumscalarsub}), and $n^{con}_\zeta$ and $n^{inf}_\zeta$ refers to Eqs. (\ref{indexscalarsup}) and (\ref{indexscalarsub}) respectively. According to this parametrization, we plot the CMB TT power spectrum of pure power-law inflation model and our model in Fig. \ref{TT}. From the figure we can see that, in most of the plotted region ($l\gtrsim 10$) our model is identified with the inflation model, and fits the data very well. This is actually not strange, because the perturbations of observable scales exits the horizon during inflation era, and should carry information of inflation part. The tiny errorbars of the data in this region also indicate that the measurement has been done accurate enough, and the shape is almost confirmed and hardly to have other alternatives. However, the data at very large scales has large error bars, and the data is not so accurate. Moreover, the data points mildly favor that there is suppression in this region. In pure power-law inflation models where all the fluctuation modes are scale-invariant, it is difficult to obtain such an suppression, however, in our models where the fluctuation modes exit the horizon before bounce, suppression takes place due to the blue-tilt of the spectrum. This result is consistent with \cite{Piao:2003zm,Cai:2015nya}, which could be seen as an important distinct from our model and inflation.
%Moreover, the modes with smaller wavenumber $k<k_0$ hasn't enter the horizon yet, so its prediction of blue-tilted spectrum will be tested in future observations. However, actually we have already got some hints in favor of such a $k$-dependence, coming from the TT spectrum of CMB which is suppressed in small $l$ region ($l<10$) \cite{Ade:2013zuv}. Since the TT spectrum corresponds to the scalar perturbations, its suppression may be explained as the blue-tilt of the perturbations in pre-inflation evolution \cite{Piao:2003zm}. According to the phenomenon, the pivot scale $k_0\sim k(l\simeq 10)\simeq 0.001\text{Mpc}^{-1}$, and since the fluctuation modes with wavenumber $k_0$ can be viewed as the last mode that exits horizon during contracting phase, it corresponds to the comoving Hubble scale of bouncing, $a_{B-}H_{B-}$. This could furtherly constrain the energy scale of the bounce \cite{Cai:2007zv}.

\section{Conclusion}
Inflation theories, although has achieved great accomplishments, has actually been suffering from the singularity problem. Phenomenologically, considering a non-singular bounce happens before inflation can help avoid the singularity. The bounce indicates that prior to our expanding universe, the universe might have been contracting from a large volume. The bounce scenario not only provide an answer to the mysterious question of ``what is our Universe like even before inflation", but more scientifically, may bring some new features to the early stage of our Universe, and provide predictions that can be discovered in the future observations.

However, to add a bounce before inflation, one should care about whether it could bring new problems. To our knowledge, the ghost instability and the cosmic anisotropy problems are among the most serious problems of the bounce scenario. One is due to the NEC violation which is necessary for the bounce, and the other is due to the growth of anisotropy during contraction. In this paper, we present a model making use of the Galileon theory, which can violate the NEC and realize bounce behavior without introducing ghosts, because of the delicate design of the Lagrangian. Moreover, a negative potential in the contracting phase, which causes the EoS of this model large than unity, is useful to avoid the dominance of the cosmic anisotropy. In order to enter into the usual inflationary era after the bounce, the potential of the model will have to be connected to some positive value functions, which can be realized phenomenologically by shape functions.

%where inflation is preceded by a non-singular bounce, solving the theoretical problems and meeting with the observational data simultaneously. In our scenario, our universe starts with a slowly-contracting phase with EoS $w\geq 1$, such that the initial anisotropies, if exist, will not go too fast so as to dominate over the background, and make the universe collapse into a totally anisotropic one. This generally requires a negative potential of the cosmic field in the contracting phase (or kinetic term much larger than potential in $w=1$ case). After contracting, the universe bounces into an expanding phase with a positive flat potential, driving ordinary inflation. The bounce can be realized by introducing the ``Galileon-term" in the field lagrangian, such that when bounce violates NEC, ghost problem will not appear.

Since now there are multi-stages in the evolution of the early universe, primordial perturbations can be generated in either in contracting phase or expanding phase. Those generated in expanding phase corresponds to large wavenumber $k$ (small scale), which has entered into horizon and can be observed by nowadays observations. Taking proper inflation potential, it is not difficult to get nearly scale-invariant $n_s$ with proper tilt, which is consistent with the PLANCK data. Meanwhile, those generated in contracting phase corresponds to small $k$ (large scale), which reenter into horizon much later, and can only be observed by future experiments. Due to the large background EoS in the contracting phase, the spectrum of these modes are blue-tilted, which will be more and more suppressed when the scales get larger and larger. However, there are already some hints in nowadays observation, i.e., the small $l$ suppression observed at TT spectra of CMB map. Moreover, the spectrum might present some oscillating features due to the bounce effect, and this can be viewed as a distinction to pure inflation models. Tensor perturbations and tensor spectrum are also discussed in this paper, which has similar behaviors as the scalar ones.

Nowadays along with the rapid development of the astrophysical observations, the observable range are getting more and more enlarged, and the observational data are also more and more accurate. More and more interesting new phenomena are discovered, which is not only chance but also challenge to the early universe model buildings. For instance, the newly released PLANCK data has pointed out that there have been several anomalies in the CMB sky, which has been stimulating great interests among astrophysicists and cosmologists. Can the bounce inflation scenario be good enough to explain those phenomena? Or what more phenomena can it predict that could be discovered by the future observations? We are trying to answer these questions in our upcoming works.

\begin{acknowledgments}
We thank Jean-Luc Lehners and Yun-Song Piao for useful discussion. The work of T.Q. is supported in part by the Open Project Program of State Key Laboratory of Theoretical Physics, Institute of Theoretical Physics, Chinese Academy of Sciences, China (No.Y4KF131CJ1), and in part by NSFC under Grant No: 11405069. The work of Y.T.W. is supported in part by NSFC under Grant No:11222546, in part by National Basic Research Program of China, No:2010CB832804.
\end{acknowledgments}

%%%%%%%%%%%%%%%%%%%%%%%%%%%%%%%%%%%%%%%%%%%%%%%%%

\end{document}